\documentclass{jfm}
\usepackage{graphicx}
\usepackage{epstopdf, epsfig}
\graphicspath{{./}}

\usepackage{subcaption}
\usepackage{hyperref}
\usepackage{color}
\newcommand{\aleck}{\textcolor{black}}

\newcommand{\bernhard}{\textcolor{black}}
\newcommand{\Rone}{\textcolor{black}}
\newcommand{\Rtwo}{\textcolor{black}}
\definecolor{orange}{rgb}{1,0.5,0}
\newcommand{\Rthree}{\textcolor{black}}
\usepackage[colorinlistoftodos, textwidth=1.5cm]{todonotes}
\usepackage{soul}

\shorttitle{Flocculation in homogeneous isotropic turbulence}
\shortauthor{K. Zhao, F. Pomes, B. Vowinckel, T.-J. Hsu, B. Bai and E. Meiburg}

\title{Flocculation of suspended cohesive particles in homogeneous isotropic turbulence}

\author{K. Zhao\aff{1,2},
  F.Pomes\aff{1},
  B. Vowinckel\aff{1,3},
  T.-J. Hsu\aff{4},
  B. Bai\aff{2},
 \and E. Meiburg\aff{1}\corresp{\email{meiburg@engineering.ucsb.edu}}
 }

\affiliation{\aff{1}Department of Mechanical Engineering, UC Santa Barbara, Santa Barbara, CA 93106, USA
\aff{2}State Key Laboratory of Multiphase Flow in Power Engineering, Xi'an Jiaotong University, Xi'an 710049, China
\aff{3}Leichtwei\ss-Institute for Hydraulic Engineering and Water Resources, Technische Universit\"{a}t Braunschweig, 38106 Braunschweig, Germany
\aff{4}Center for Applied Coastal Research, Department of Civil \& Environmental Engineering, University of Delaware, Newark, DE 19716, USA
}

\begin{document}

\maketitle

\begin{abstract}

We investigate the dynamics of cohesive particles in homogeneous isotropic turbulence, based on one-way coupled simulations that include Stokes drag, lubrication, cohesive and direct contact forces. We observe a transient flocculation phase characterized by a growing average floc size, followed by a statistically steady equilibrium phase. We analyze the temporal evolution of floc size and shape due to aggregation, breakage, and deformation. Larger turbulent shear and weaker cohesive forces yield \Rone{elongated flocs that are smaller in size}. Flocculation proceeds most rapidly when the fluid and particle time scales are balanced and a suitably defined Stokes number is \textit{O}(1). During the transient stage, cohesive forces of intermediate strength produce \Rone{flocs of the largest size}, as they are strong enough to cause aggregation, but not so strong as to pull the floc into a compact shape. Small Stokes numbers and weak turbulence delay the onset of the equilibrium stage. During equilibrium, stronger cohesive forces yield \Rone{flocs of larger size}. The equilibrium floc size distribution exhibits a preferred size that depends on the cohesive number. We observe that flocs are generally elongated by turbulent stresses before breakage. Flocs of size close to the Kolmogorov length scale preferentially align themselves with the intermediate strain direction and the vorticity vector. \Rone{Flocs of smaller size} tend to align themselves with the extensional strain direction. More generally, flocs are aligned with the strongest Lagrangian stretching direction. The Kolmogorov scale is seen to limit floc growth. We propose a new flocculation model with a variable fractal dimension that predicts the temporal evolution of the floc size and shape.
\end{abstract}

\begin{center}
------------------------------------------------------------------------------------------------------------------
\end{center}


\section{Introduction}

Individual cohesive particles suspended in liquid or gaseous fluid flows tend to form larger aggregates, due to attractive inter-particle forces that cause the primary particles to flocculate. This mechanism plays a dominant role in environmental processes such as sediment erosion and transport in rivers and oceans, or soil erosion by wind \citep{winterwerp2002flocculation, guo2011freshwater, wang2013sediment, tarpley2019tidal}. In planetary astrophysics, corresponding processes influence the coagulation of dust during the formation of protoplanetary disks \citep{ormel2007dust, schafer2007collisions, ormel2009dust, ormel2011dust}. The emergence of large aggregates due to the flocculation of cohesive primary particles is also highly relevant in the context of a wide range of industrial processes, such as the ingestion of dust in gas turbine engines \citep{bons2017simple, sacco2018dynamic}, or the use of membrane separation technologies for wastewater treatment and the production of potable water \citep{bratskaya2006enhanced, leiknes2009effect, moghaddam2010coagulation, kang2012flocculation}. Similarly, the operation of certain types of medical equipment, for example dry powder inhalers \citep{yang2013analysis, yang2015numerical, tong2013cfd, tong2016cfd}, involves the formation of agglomerates or flocs. The flocculation process is strongly affected by the turbulent nature of the underlying fluid flow. Small-scale eddies modify the collision dynamics of the primary particles and hence the growth rate of the flocs, while turbulent stresses can result in the deformation and breakup of larger cohesive flocs. Hence the dynamic equilibrium between floc growth and breakup is governed by a complex and delicate balance of hydrodynamic and inter-particle forces.

A host of experimental studies have provided considerable insight into key aspects of the development of flocs in turbulent shear flows, such as their growth rate \citep{biggs2000activated, yu2006flocculation, xiao2010comparative, kuprenas2018shear}, the equilibrium size distribution \citep{chaignon2002evolution, bouyer2004experimental, rahmani2004evolution, lee2020estimation}, and the transient shape of the flocs \citep{maggi2007effect, he2012characteristic, guerin2017dynamics}. Based on the early pioneering work by \citet{levich1962physicochemical}, several of these investigations have employed a population balance approach to formulate models for the temporal floc evolution \citep{maggi2007effect, shin2015stochastic, winterwerp1998simple, son2008flocculation, son2009effect}. Alternative approaches based on the classical work by \citet{smoluchowski1936three} propose statistical collision equations \citep{ives1973theory, yang2013new, klassen2017three}. Most of the above approaches do not incorporate detailed information on the overall floc strength, which varies with the floc size and shape, and with the strength of the bonds between the primary cohesive particles \citep{dizaji2019collision}. \citet{moreno2006mechanistic}, on the other hand, consider the dependence of the overall floc strength on the number and strength of the bonds within the floc. \citet{nguyen2014numerical} and \citet{gunkelmann2016influence} observe that loosely structured agglomerates fragment more easily during collisions than densely packed ones.

In recent years, highly resolved numerical simulations have begun to provide a promising new avenue for gaining insight into the interplay of hydrodynamic, inertial and inter-particle forces during the growth, deformation and breakup of aggregates \citep{marshall2014adhesive}. The study by \cite{zhao2020efficient} focuses on a conceptually simple cellular model flow in order to explore the competition between inertial, drag, and cohesive forces during the flocculation process. The authors find that floc growth proceeds most rapidly if the fluid and particle time scales are in equilibrium, so that a suitably defined Stokes number is of order unity. Based on simulations in a similar model flow, \cite{ruan2020structural} suggest a criterion for the breakup of aggregates. \cite{dizaji2019collision} investigate the dynamics, collision and fragmentation of flocs in shear flows, via two-way coupled simulations that account for the modification of the flow by the particles. They demonstrate that the particle-fluid interaction induces vortex rings in the flow. \cite{dizaji2016accelerated} propose a novel stochastic vortex structure method, and proceed to show that this numerical approach produces realistic collision rates in homogeneous turbulence. For flocculation in turbulence, \cite{dizaji2017significance} show that the aggregation process influences the background turbulence only weakly. Quite recently, \citet{chen2019exponential} and \citet{chen2020collision} conducted a detailed computational study of cohesive particle aggregation in homogeneous isotropic turbulence, based on two-way coupled direct numerical simulations combined with an adhesive discrete element method. The simulations presented in \citet{chen2019exponential}, which account for Stokes drag, lubrication and adhesive contact forces, address the early stages of flocculation before an equilibrium size distribution is reached. Upon the onset of flocculation, the results demonstrate a time-dependent, exponential size distribution of the flocs for all values of the cohesive force strength. Based on this observation, the authors develop an effective agglomeration kernel for the population balance equation that successfully reproduces the DNS results. In a follow-up study, \citet{chen2020collision} investigate the collision-induced breakup of agglomerates in homogeneous isotropic turbulence. The authors are able to quantify the fraction of collisions that result in breakage, which presents useful information for closing the population balance equation. However, because the simulations focus on the early stages of flocculation before the emergence of an equilibrium size distribution, and because they employ particles with diameter approximately equal to the Kolmogrov scale, they do \Rone{not} allow the authors to assess the role of the Kolmogorov length scale in limiting the floc size, a widely reported experimental observation \citep{fettweis2006suspended, coufort2008analysis, braithwaite2012controls, kuprenas2018shear}. Furthermore, the authors model the cohesive van der Waals force as a ``sticky force" that acts only on contact. Several previous studies, on the other hand, have indicated that this attractive force extends over a finite range even before the particles come into contact, so that it can affect the probability that two close-by particles will collide \citep{visser1989van, israelachvili1992adhesion, wu2017aggregation, vowinckel2019settling}. 

The present investigation aims to explore the interplay between floc aggregation, deformation and breakup from inception all the way to the dynamic equilibrium phase, with the goal of obtaining scaling laws for both of these qualitatively different stages. Towards this end, we will employ a simulation approach that tracks dispersed individual spherical particles of a given diameter in homogeneous isotropic turbulence. The simulations are one-way coupled in the sense that the particles do not modify the fluid flow, although particle-particle interactions are fully accounted for, and the grid spacing employed for calculating the fluid motion is smaller than the particle diameter. Sometimes this approach is referred to as ``three-way coupled''. The simulations account for inter-particle forces based on recently developed advanced collision models for viscous flows 
\citep[][and references therein]{biegert2017collision},
along with the cohesive force model of \cite{vowinckel2019settling}. The homogeneous isotropic turbulence is generated and maintained via the forcing method of \citet{eswaran1988examination}. We will employ these simulations in order to investigate the floc size and shape evolution, the floc size distribution during the equilibrium stage, the orientation of the flocs with regard to the principal directions of the Eulerian strain and the Lagrangian stretching, as well as the role of the Kolmogorov length scale in limiting floc growth. Based on our findings, we then propose a novel flocculation model that predicts the evolution of the floc size and shape with time. To assess the performance of this new flocculation model, we will compare its predictions to those obtained with existing models in the literature.

The paper is structured along the following lines. Section \ref{sec:Numerical method} briefly reviews the governing equations for the fluid flow and the particle motion, and it describes the computational approach. It identifies the governing dimensionless parameters and quantifies the range over which they will be varied in the present investigation. The properties of the turbulent flow fields are described in Section \ref{sec:Simulation of single-phase turbulence}, and their statistically stationary and isotropic nature is discussed. Starting from 10,000 randomly distributed individual particles, we then analyze the temporal evolution of the floc size and shape as a result of aggregation, deformation and breakage in Section \ref{sec:Flocculation of cohesive particles}. Here we will distinguish between the transient flocculation stage and the equilibrium stage, and we will discuss the underlying physical mechanisms. We will furthermore analyze the alignment of the flocs with regard to the principal strain directions of the turbulent velocity field, and we will focus on how the Kolmogorov scale affects the maximum floc size. Subsequently, we introduce the new flocculation model in Section \ref{sec:A new flocculation model with variable fractal dimension}, and we compare its predictions to those obtained from existing models. Section \ref{sec:Conclusions} summarizes the main findings of the current investigation, and presents its key conclusions.

\section{Governing equations and numerical method}\label{sec:Numerical method}
\subsection{Particle motion in homogeneous isotropic turbulence} \label{subsec:Particle motion in homogeneous-isotropic turbulence}

We consider the one-way coupled motion of suspended cohesive particles in three-dimensional, incompressible homogeneous isotropic turbulence. The motion of the single-phase fluid with constant density $\rho_f$ and kinematic viscosity $\nu$ is governed by

\begin{equation} \label{eq:fluid mass conservation}
    \nabla \cdot \boldsymbol{u_f} = 0 \ ,
\end{equation} 
\begin{equation} \label{eq:fluid momentum conservation}
    \frac{\partial \boldsymbol{u_f} }{\partial  t} + (\boldsymbol{u_f} \cdot \nabla) \boldsymbol{u_f} = - \frac{1}{\rho_f} \nabla p + \nu \nabla^2 \boldsymbol{u_f} + \boldsymbol F_{tur} \ ,
\end{equation} 
where $\boldsymbol{u_f} = (u_f, v_f, w_f)^{\rm T}$ denotes the fluid velocity vector and $p$ indicates the hydrodynamic pressure. \Rone{We employ the spectral approach of \citet{eswaran1988examination} to obtain the forcing term $\boldsymbol F_{tur}$, which generates and maintains statistically stationary turbulence, as implemented in \citet{chouippe2015forcing}.} Here, $\boldsymbol F_{tur}$ is non-zero only in the low-wavenumber band where the wavenumber vector $|\boldsymbol \kappa| < \kappa_f$, with $\kappa_f = 2.3\kappa_0$ and $\kappa_0 = 2 \pi / L_0$, with $L_0$ denoting the length of the physical domain. The origin $\boldsymbol \kappa = 0$ is not forced. In addition to the cutoff wavenumber $\kappa_f$, the random forcing process is governed by the dimensionless parameter $D_s = \sigma^2 T_0 L_0^4 / \nu^3$, where $\sigma^2$ and $T_0$ indicate the variance and the time scale of the random process, respectively. Regarding the details of evaluating $\boldsymbol F_{tur}$ from $\kappa_f$ and $D_s$, we refer the reader to the original work by \citet{eswaran1988examination}.  

We approximate each primary suspended particle $i$ as a sphere moving with translational velocity ${\boldsymbol u}_{p,i} = (u_{p,i}, v_{p,i}, w_{p,i})^{\rm T}$ and angular velocity ${\boldsymbol \omega}_{p,i}$. These are obtained from the linear and angular momentum equations
\begin{equation} \label{eq:particle motion}
    m_p \frac{\mathrm{d}{\boldsymbol u}_{p,i}}{\mathrm{d}t} = {\boldsymbol F}_{d,i} + \underbrace{\sum_{j=1,j \ne i}^{N}(\boldsymbol F_{con,ij} + \boldsymbol F_{lub,ij} + \boldsymbol F_{coh,ij})}_{{\boldsymbol F}_{c,i}} \ ,
\end{equation}
\begin{equation} \label{eq:particle rotation}
    I_p \frac{\mathrm{d}{\boldsymbol \omega_{p,i}}}{\mathrm{d}t} = \underbrace{\sum_{j=1,j \ne i}^{N}(\boldsymbol T_{con,ij} + \boldsymbol T_{lub,ij})}_{{\boldsymbol T}_{c,i}} \ ,
\end{equation}
where the primary particle $i$ moves in response to the Stokes drag force $\boldsymbol F_{d,i} = -3 \pi D_p \mu ({\boldsymbol u}_{p,i} - {\boldsymbol u}_{f,i})$, and the particle-particle interaction force $\boldsymbol F_{c,i}$. Buoyancy is not considered here, so that we can investigate the effects of particle inertia in isolation. \Rtwo{We only consider primary particles that are larger than 2$\mu m$ (cf. table 1), so that \bernhard{a suitably defined Peclet number measuring the relative importance of hydrodynamic and Brownian forces is sufficiently large for their Brownian motion to be negligible \citep{biegert2017collision, chen2019exponential, vowinckel2019settling, partheniades2009}.}} ${\boldsymbol u}_{p,i}$ indicates the particle velocity evaluated at the particle center. ${\boldsymbol u}_{f,i} = \sum_{1}^{N_i}(\phi_{i,k} {\boldsymbol u}_{f,k})$ represents the fluid velocity averaged over the volume of particle $i$, where $N_{i}$ denotes the number of Eulerian grid cells covered by particle $i$, ${\boldsymbol u}_{f,k}$ is the fluid velocity at the center of the grid cell $k$, and $\phi_{i,k}$ is the volume fraction of the particle $i$ in the grid cell $k$. We remark that the above implies that the diameter $D_p$ of the primary particle should be larger than the grid spacing $h$. This avoids the need for interpolating the fluid velocity within one grid cell, which would be required if $D_p < h$ \citep{chen2019exponential}. $m_p$ denotes the particle mass, $\mu$ the dynamic viscosity of the fluid, and $N$ the total number of particles in the flow. We assume all particles to have the same diameter $D_p$ and density $\rho_p$. $\boldsymbol F_{c,i}$ accounts for the direct contact force $\boldsymbol F_{con,ij}$ in both the normal and tangential direction, as well as for short-range normal and tangential forces due to lubrication $\boldsymbol F_{lub,ij}$ and cohesion $\boldsymbol F_{coh,ij}$, where the subscript $ij$ indicates the interaction between particles $i$ and $j$. $I_p = \pi \rho_p D_p^5 / 60$ denotes the moment of inertia of the particle. $\boldsymbol T_{c,i}$ represents the torque due to particle-particle interactions, where we distinguish between direct contact torque $\boldsymbol T_{con,ij}$ and lubrication torque $\boldsymbol T_{lub,ij}$. \Rtwo{Within a large floc, we account for all of the individual binary particle interactions.}

The lubrication force $\boldsymbol F_{lub,ij}$ is accounted for based on \citet{cox1967slow} as implemented in \citet{zhao2020efficient}. \bernhard{We note that, although the present study is limited to monodisperse particles, polydisperse particle-particle interactions can be taken into account by an effective radius $R_{eff}=R_p R_q / (R_p + R_q)$, where $R_p$ and $R_q$ are the radii of two interacting spheres.} \Rtwo{Following \citet{biegert2017collision}, the collision force $\boldsymbol F_{con,ij}$ is represented by a nonlinear spring–dashpot model in the normal direction,} \Rthree{while the tangential component is modelled by a linear spring–dashpot model capped by the Coulomb friction law to account for zero-slip rolling or sliding of particles. We note that the tangential component of the contact force depends on the surface roughness, a prescribed restitution coefficient $e_{dry} = 0.97$ and a friction coefficient $e_{fri} = 0.15$ are implemented to yield adaptively calibration for every collision as described by \citet{biegert2017collision}.} \Rtwo{The cohesive force $\boldsymbol F_{coh,ij}$, which reflects the combined influence of the attractive van der Waals force and the repulsive electrostatic force, is based on the work of \citet{vowinckel2019settling}, where additional details and validation results are provided. The model assumes a parabolic force profile, distributed over a thin shell surrounding each primary particle.} \Rtwo{Hence the cohesive force between primary particles extends over a finite range, so that it is felt by the particles even before they come into direct contact. We consider two primary particles to be part of the same floc when their surface distance is smaller than half the range of the cohesive force, as implemented in \citet{zhao2020efficient}.} \Rthree{We remark that, based on equations (\ref{eq:particle motion}) and (\ref{eq:particle rotation}), the configuration of the primary particles within a floc can change with time in response to fluid forces, since the cohesive bonds are not rigid. Specifically, the contact points on the surface of the primary particles are not fixed, so that the primary particles can rotate individually within a floc.}

\subsection{Nondimensionalization} \label{subsec:Non-dimensionalization}

In order to render the above governing equations dimensionless, we consider primary particles with diameter $D_p = 5 \ \rm{\mu m}$, which represents a typical value for clay or fine silt. The cubic computational domain has an edge length $L_0 = 125D_p = 6.25 \times 10^{-4} \ \rm{m}$. As time scale of the random turbulent forcing process we select $T_0 = 7.81 \times 10^{-5} \ \rm{s}$. By choosing $L_0$, $T_0$ and $\rho_f = 1,000 \ \rm{kg/m^{3}}$ as the characteristic length, time and density scales, we obtain the characteristic velocity scale $U_0 = L_0/T_0 = 8 \ \rm{m/s}$, which is similar to values employed in previous investigations \citep{chen2019exponential, chen2020collision}. We employ $L_0$ and $U_0$ to define the turbulence Reynolds number $Re = L_0 U_0 \rho_f/ \mu$. 

The dimensionless continuity and momentum conservation equations can then be expressed as
\begin{equation} \label{eq:dimensionless fluid mass conservation}
    \tilde {\nabla} \cdot \tilde {\boldsymbol u}_f = 0 \ ,
\end{equation} 
\begin{equation} \label{eq:dimensionless fluid momentum conservation}
    \frac{\partial \tilde {\boldsymbol u}_f }{\partial \tilde t} + (\tilde {\boldsymbol u}_f \cdot \tilde {\nabla}) \tilde {\boldsymbol u}_f = - \tilde {\nabla} \tilde p + \frac{1}{Re} \tilde {\nabla}^2 \tilde {\boldsymbol u}_f + \tilde {\boldsymbol F}_{tur} \ ,
\end{equation} 
while the dimensionless equations of motion for the primary cohesive particles take the form
\begin{equation} \label{eq:dimensionless particle motion}
    \tilde m_p \frac{\mathrm{d} \tilde {\boldsymbol u}_{p,i}}{\mathrm{d} \tilde t}  =  \underbrace{- \frac {3 \pi \tilde D_p (\tilde {\boldsymbol u}_{p,i} - \tilde {\boldsymbol u}_{f,i})}{Re}}_{\tilde {\boldsymbol F}_{d,i}} + \sum_{j=1,j \ne i}^{N}(\tilde {\boldsymbol F}_{con,ij} + \tilde {\boldsymbol F}_{lub,ij} + \tilde {\boldsymbol F}_{coh,ij} ) \ ,
\end{equation}
\begin{equation}
    \tilde I_p \frac{\mathrm{d} \tilde {\boldsymbol \omega}_{p,i}}{\mathrm{d} \tilde t} = \sum_{j=1,j \ne i}^{N}(\tilde {\boldsymbol T}_{con,ij} + \tilde {\boldsymbol T}_{lub,ij}) \ .
\end{equation}
Here dimensionless quantities are denoted by a tilde. The dimensionless particle mass is defined as $\tilde m_p = \pi \tilde D_p^3 \tilde \rho_s / 6$, the moment of inertia $\tilde I_p = \pi \tilde \rho_s \tilde D_p^5 / 60$, and the density ratio $\tilde \rho_s = \rho_p / \rho_f$. The dimensionless direct contact and lubrication forces, $\tilde {\boldsymbol F}_{con,ij}$ and $\tilde {\boldsymbol F}_{lub,ij}$, are accounted for based on \citet{zhao2020efficient}, while the dimensionless cohesive force $\tilde {\boldsymbol F}_{coh,ij}$ is defined as
\begin{equation}
    \tilde {\boldsymbol F}_{coh,ij} = \left\{
    \begin{array}{ll}
        - 4 Co \frac {\tilde \zeta_{n,ij}^2 - \tilde h_{co} \tilde \zeta_{n,ij}}{\tilde h_{co}^2} \boldsymbol n,        &  \tilde \zeta_{min} < \tilde \zeta_{n,ij} \leqslant \tilde h_{co} \ , \\ 
        0,   & \rm{otherwise} \ .
    \end{array}\right.
\end{equation}
Here $\tilde \zeta_{min} = 0.0015 \tilde D_p$ and $\tilde h_{co} = 0.05 \tilde D_p$ represent the surface roughness of the particles and the range of the cohesive force, respectively. $\boldsymbol n$ represents the outward-pointing normal on the particle surface, while $\tilde \zeta_{n, ij}$ is the normal surface distance between particles $i$ and $j$. The cohesive number $Co$ indicates the ratio of the maximum cohesive force $\vert\vert{\boldsymbol F_{coh,ij}}\vert\vert$ at $\tilde \zeta_{n,ij} = \tilde h_{co}/2$ to the characteristic inertial force
\begin{equation} \label{eq:Co_definition}
    Co = \frac {{\rm max} (\vert\vert{\boldsymbol F_{coh,ij}}\vert\vert)}{U_0^2 L_0^2 \rho_f} = \frac {A_H D_p}{16 h_{co} \zeta_{0} } \frac {1}{U_0^2 L_0^2 \rho_f } \ ,
\end{equation}
where the Hamaker constant $A_H$ is a function of the particle and fluid properties, and the characteristic distance $\zeta_0  = 0.00025D_p$. \citet{vowinckel2019settling} provide representative values of \bernhard{various physicochemical parameters such as $A_H$, salt concentration and grain size of the primary particles} for common natural systems. The present numerical approach for simulating the dynamics of cohesive sediment has been employed to predict the flocculation in simple vortical flow fields, and it was successfully validated with experimental data in our earlier work \citep{zhao2020efficient}.

To summarize, the simulations require as direct input parameters the turbulence Reynolds number $Re$, the characteristic parameter of the random turbulent forcing process $D_s$, the dimensionless particle diameter $\tilde D_p$, the total number of particles $N$, the density ratio $\tilde \rho_s$, and the cohesive number $Co$. As we will discuss below, $Re$ and $D_s$ can equivalently be expressed by the shear rate $G$ of the turbulence, cf. equation (\ref{eq:defination_G}), and the Stokes number $St$ defined by equation (\ref{eq:defination_St}). A list of the relevant dimensionless parameters is provided in table \ref{tab:table1}. We remark that due to computational limitations the simulations consider Kolmogorov scales that are somewhat smaller than typical field values, and turbulent shear rates that are larger than field values. Hence the ratio of the Kolmogorov length scale to the primary particle size takes values up to 3.3 in the simulations, as compared to values up to $O(10)$ under typical field conditions. For convenience, the tilde symbol will be omitted henceforth.

\begin{table}
  \begin{center}
\def~{\hphantom{0}}
  \begin{tabular}{lll}
    \textbf{Physical parameters} \\
    \\
    Particle diameter & $D_p$ & $5 \ \rm{\mu m}$   \\ 
    \\
    Particle density & $\rho_p$ & $2.65 \times 10^{3} - 5.0 \times 10^{4} \ \rm{kg/m^{3}}$   \\
    \\
    Number of particles & $N$ & $1.0 \times 10^{4}$  \\
    \\
    Volume fraction of particles & $\phi_p$ & $2.68 \times 10^{-3}$  \\
    \\
    Hamaker constant & $A_H$ & $1.0 \times 10^{-20} - 3.0 \times 10^{-18}\ \rm{J}$   \\
    \\
    Fluid density & $\rho_f$ & $1.0 \times 10^{3} \ \rm{kg/m^{3}}$  \\
    \\
    Dynamic viscosity \ \ & $\mu$ \ \ & $1.0 \times 10^{-3}$ -  $2.5 \times 10^{-3}$ \ $\rm{Pa \ s}$ \\
    \\
    Shear rate \ \ & $G$ \ \ & $3.7 \times 10^{3} -  9.5 \times 10^{4} \ \rm{s^{-1}}$ \\
    \\
    Kolmogorov length scale \ \ & $\eta$ \ \ & $3.25 - 16.5 \ \rm{\mu m}$ \\
    \\
    \textbf{Non-dimensional parameters} \\
    \\
    Particle diameter & $\tilde D_p$ & $8.0 \times 10^{-3}$   \\
    \\
    Density ratio & $\tilde \rho_s$ & 2.65 - 50   \\
    \\
    Turbulence Reynolds number & $Re$ & $2.0 \times 10^{3} - 5.0 \times 10^{3}$ \\
    \\
    Turbulent forcing parameter & $D_s$ & $1.0 \times 10^{4} - 1.0 \times 10^{7}$ \\
    \\
    Cohesive number & $Co$ & $4.0 \times 10^{-10} - 1.2 \times 10^{-7}$  \\
    \\
    Shear rate & $\tilde G$  & 0.29 -  7.4  \\
    \\
    Stokes number & $St$  & 0.02 -  1.92  \\
    \\
  \end{tabular}
  \caption{Nondimensionalization employed in the present work: the characteristic values for length, velocity and density are $L_0 = 125D_p = 6.25 \times 10^{-4} \ \rm m$, $U_0 = 8 \ \rm{m/s}$ and $\rho_f = 1,000 \ \rm{kg/m^3}$, respectively.}
  \label{tab:table1}
  \end{center}
\end{table}

\section{Simulation of single-phase turbulence} \label{sec:Simulation of single-phase turbulence}

\subsection{Computational set-up} \label{subsec:Computational set-up}

The triply periodic computational domain $\Omega$ has a dimensionless size of $L_x \times L_y \times L_z = 1 \times 1 \times 1$, with the number of grid cells $N_x \times N_y \times N_z = 128 \times 128 \times 128$. This relatively modest number of grid points enables us to conduct the simulations over sufficiently long times for the flocculation and break-up processes to reach an equilibrium state \citep{tran2018changes}, and it is in line with the earlier study of \cite{chen2019exponential}. As mentioned above, we set the diameter $D_p$ of the primary particles moderately larger than the grid size $h = L_x/N_x$, at a constant value $D_p/h=1.024$.

Before introducing the particles into the flow, we simulate the single-phase turbulence until it reaches a statistically stationary state. Table \ref{tab:single-phase turbulence simulations} gives an overview of the physical parameters for the simulations conducted within the present investigation. Here the Kolmogorov length scale and the root-mean-square velocity are defined as $\eta = 1/ (Re^3 \epsilon)^{1/4}$ and $u_{rms} = (2 k /3)^{1/2}$, respectively, where $\epsilon$ and $k$ denote the domain-averaged dissipation rate and kinetic energy of the fluctuations. The Taylor Reynolds number $Re_{\lambda} = \lambda u_{rms} Re$ of the turbulence is based on the Taylor microscale $\lambda = \sqrt{15} \, u_{rms}/(Re \, \epsilon)^{1/2}$. To provide a more complete quantitative description of the fluid shear, we define the vorticity fluctuation amplitude
\begin{equation} \label{eq:defination_G}
    G = \frac{1}{Re \, \eta^2}  \ ,
\end{equation}
which can also be regarded as the turbulent shear rate. For additional details with regard to these quantities, we refer the reader to \citet{pope2001turbulent}. 

\begin{table}
  \begin{center}
\def~{\hphantom{0}}
  \begin{tabular}{ccccccc}
    Case \ \ \ & $Re$ \ \ \ & $D_s$ \ \ \ & $Re_{\lambda}$ \ \ \ & $\eta$ \ \ \  & $u_{rms}$ \ \ \ & $G$   \\ 
    \\
    Tur1 \ \  & $5.0 \times 10^3$ & $1.0 \times 10^4$ & $9.72$ & $2.64 \times 10^{-2}$ & $1.2 \times 10^{-2}$ & $0.29$   \\ 
    \\
    Tur2 \ \ & $2.0 \times 10^3$ & $5.5 \times 10^4$ & $16.05$ & $1.83 \times 10^{-2}$ & $5.56 \times 10^{-2}$ & $1.49$   \\ 
    \\
    Tur3 \ \ & $3.0 \times 10^3$ & $5.1 \times 10^4$ & $16.97$ & $1.8 \times 10^{-2}$ & $3.88 \times 10^{-2}$ & $1.03$   \\ 
    \\
    Tur4 \ \ & $4.0 \times 10^3$ & $5.05 \times 10^4$ & $17.25$ & $1.79 \times 10^{-2}$ & $2.94 \times 10^{-2}$ & $0.78$   \\ 
    \\
    Tur5 \ \ & $5.0 \times 10^3$ & $5.0 \times 10^4$ & $15.21$ & $1.8 \times 10^{-2}$ & $2.2 \times 10^{-2}$ & $0.62$   \\ 
    \\
    Tur6 \ \ & $5.0 \times 10^3$ & $1.0 \times 10^5$ & $21.83$ & $1.48 \times 10^{-2}$ & $3.2 \times 10^{-2}$ & $0.91$   \\ 
    \\
    Tur7 \ \ & $5.0 \times 10^3$ & $1.0 \times 10^6$ & $34.65$ & $8.6 \times 10^{-3}$ & $6.88 \times 10^{-2}$ & $2.7$   \\ 
    \\
    Tur8 \ \ & $5.0 \times 10^3$ & $1.0 \times 10^7$ & $50.34$ & $5.2 \times 10^{-3}$ & $1.38 \times 10^{-1}$ & $7.4$   \\ 
    \\
  \end{tabular}
  \caption{Physical parameters of the single-phase turbulence simulations. As input parameters we specify the fluid Reynolds number $Re = L_0 U_0 / \nu$ and the characteristic parameter of the random turbulent forcing process $D_s = \sigma^2 T_0 L_0^4 Re^3$. The simulation then yields the Taylor Reynolds number $Re_{\lambda} = \lambda u_{rms} Re$, the Kolmogorov scale $\eta$, the average root-mean-square velocity $u_{rms}$, and the shear rate $G = 1/ (Re \, \eta^2)$. All of these output quantities are obtained by averaging over space and time, after a statistically stationary state has evolved.}
  \label{tab:single-phase turbulence simulations}
  \end{center}
\end{table}

\subsection{Turbulence properties for different $Re_{\lambda}$} \label{subsec:Simulation conditions}

One key goal of the present investigation is to study the flocculation of primary particles whose diameter $D_p$ is smaller than the Kolmogorov length scale $\eta$. Since the particle diameter needs to be larger than the grid spacing, and the number of grid points is limited, suitable values of $\eta$ require a relatively low $Re_{\lambda}$. On the other hand, it is known that for $Re_{\lambda} \leq O(50)$ the turbulence may not be fully developed and isotropic \citep{mansour1994decay}. Hence this section presents a more detailed discussion of the turbulence properties for $Re_{\lambda} \leq O(50)$.

\begin{figure}
    \begin{subfigure}{0.5\textwidth}
    \centering
    \caption{}
    \includegraphics[width=0.9\textwidth]{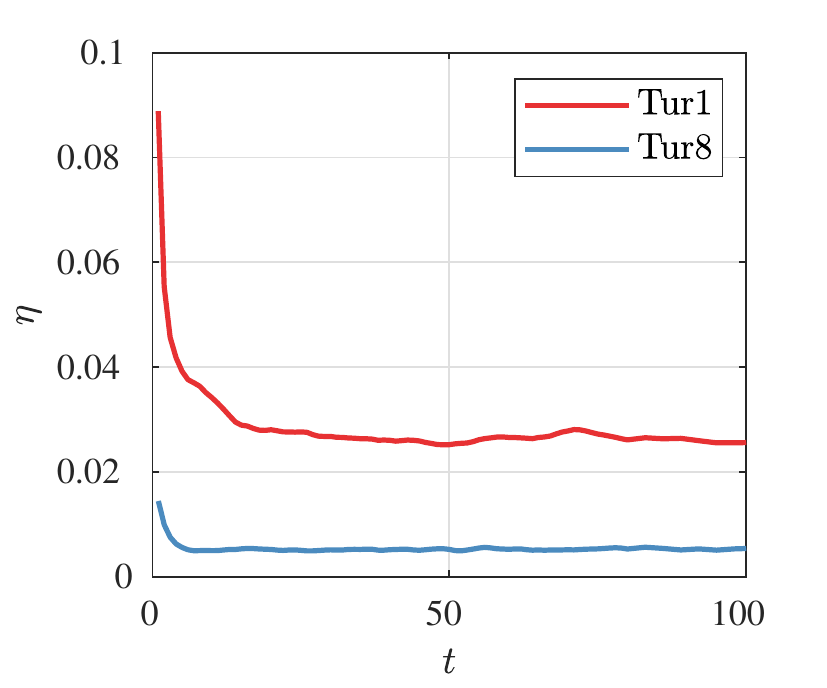}
    \label{fig:eta_t}
    \end{subfigure}
    \begin{subfigure}{0.5\textwidth}
    \centering
    \caption{}
    \includegraphics[width=0.9\textwidth]{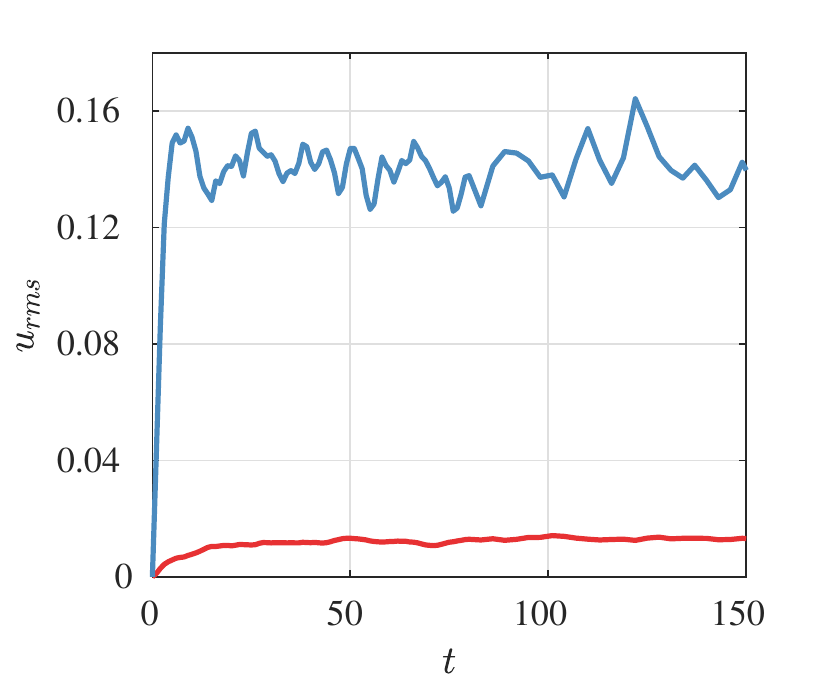}
    \label{fig:urms_t}
    \end{subfigure}
    \begin{subfigure}{0.5\textwidth}
    \centering
    \caption{}
    \includegraphics[width=0.9\textwidth]{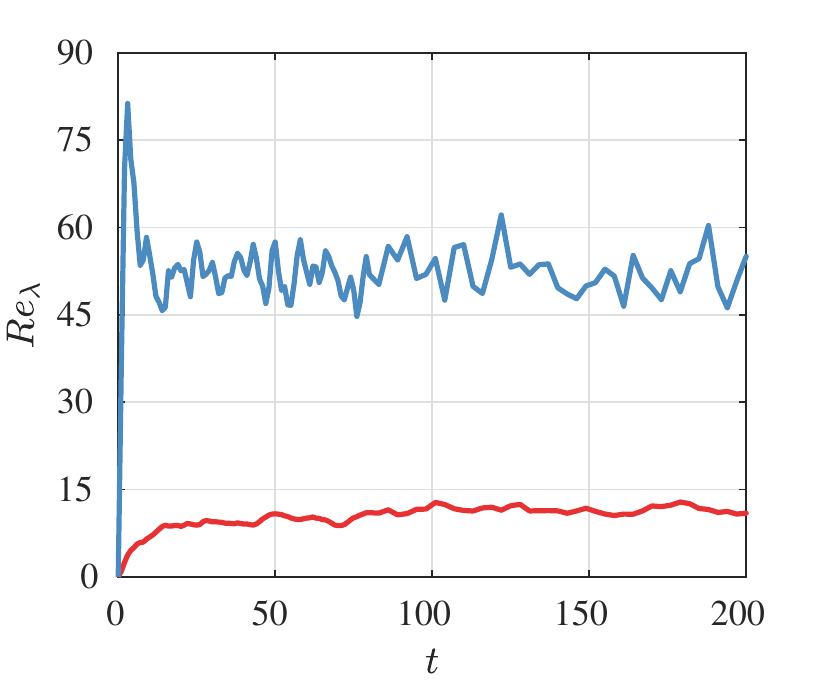}
    \label{fig:Relambda_t}
    \end{subfigure}
    \begin{subfigure}{0.5\textwidth}
    \centering
    \caption{}
    \includegraphics[width=0.9\textwidth]{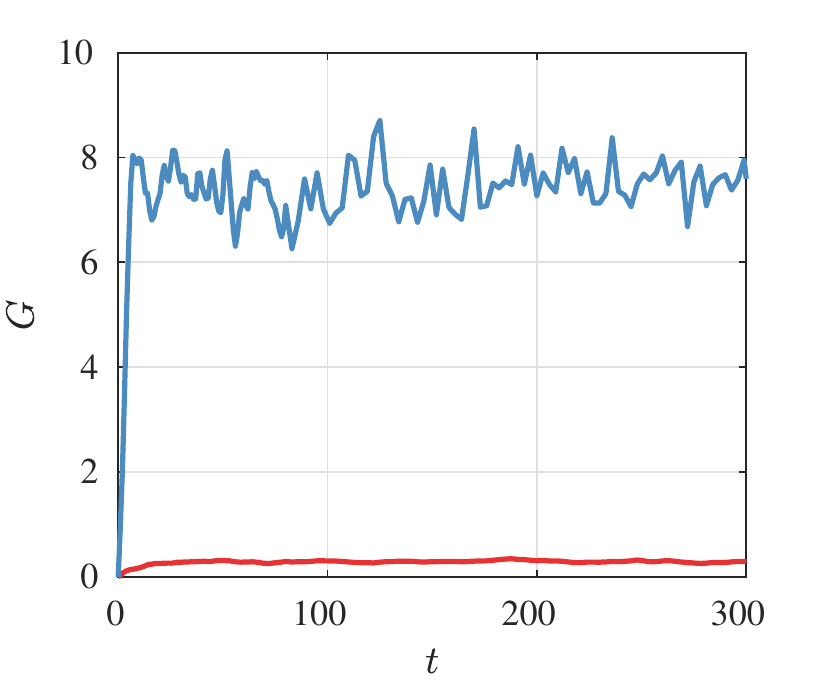}
    \label{fig:G_t}
    \end{subfigure}
    \caption{Temporal evolution of box-averaged turbulence properties for cases Tur1 and Tur8 in table \ref{tab:single-phase turbulence simulations}: (a) Kolmogorov length scale $\eta$; (b) Root-mean-square velocity $u_{rms}$; (c) Taylor Reynolds number $Re_{\lambda}$; (d) Shear rate $G$. A statistically stationary state is seen to evolve for all quantities.}
    \label{fig:temporal evolution of box-averaged quantities}
\end{figure}

\begin{figure}
    \begin{subfigure}{0.5\textwidth}
    \centering
    \caption{}
    \includegraphics[width=0.9\textwidth]{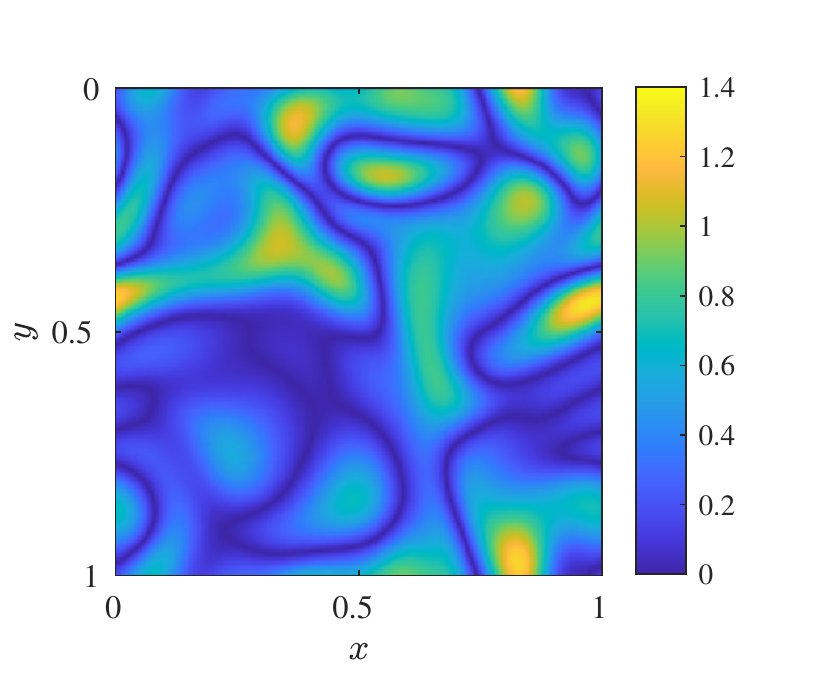}
    \label{fig:vorticity_Tur1}
    \end{subfigure}
    \begin{subfigure}{0.5\textwidth}
    \centering
    \caption{}
    \includegraphics[width=0.9\textwidth]{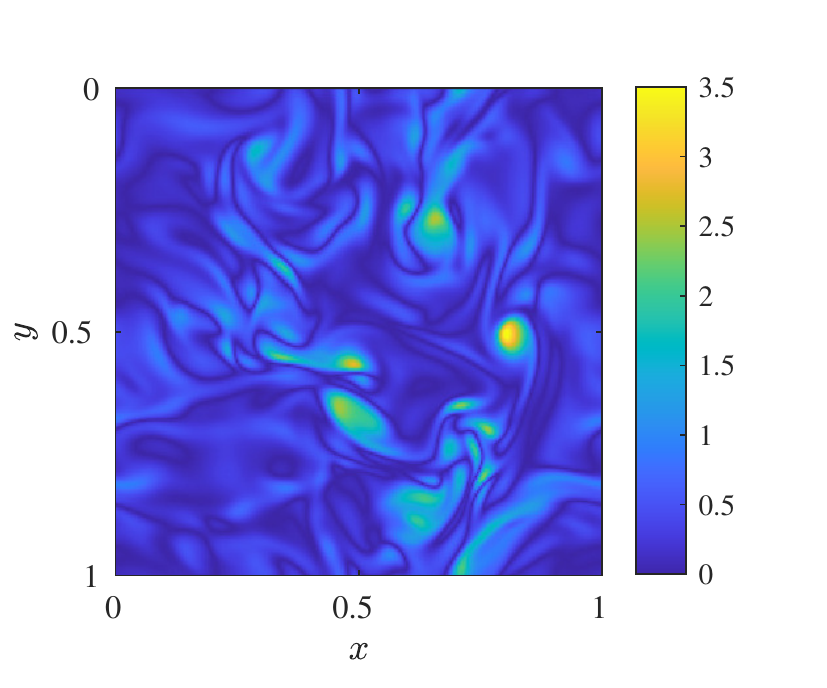}
    \label{fig:vorticity_Tur8}
    \end{subfigure}
    \caption{Representative snapshots of the vorticity modulus normalized by the vorticity fluctuation amplitude $G$, shown in the plane $z = 0.5$. (a) Case Tur1; (b) Case Tur8.}
    \label{fig:the modulus of vorticity}
\end{figure}

Figure \ref{fig:temporal evolution of box-averaged quantities} shows the time-dependent evolution of the box-averaged Kolmogorov length $\eta$, the root-mean-square velocity $u_{rms}$, the Taylor Reynolds number $Re_{\lambda}$, and the shear rate $G$ for cases Tur1 and Tur8, which have time-averaged Taylor Reynolds numbers of 9.72 and 50.34, respectively. Both cases are seen to reach statistically stationary states. We note that while case Tur8 results in $\eta/h = 0.6656$, \citet{chouippe2015forcing} demonstrated the validity of the current turbulent forcing approach even when the Kolmogorov length is smaller than the grid spacing. Snapshots of the vorticity modulus in a slice of the computational domain are shown in Figure \ref{fig:the modulus of vorticity}. They exhibit the intermittent multiscale patterns featuring eddies of different size along with thin filaments that are typical for turbulence.

\begin{figure}
    \begin{subfigure}{0.5\textwidth}
    \centering
    \caption{}
    \includegraphics[width=0.9\textwidth]{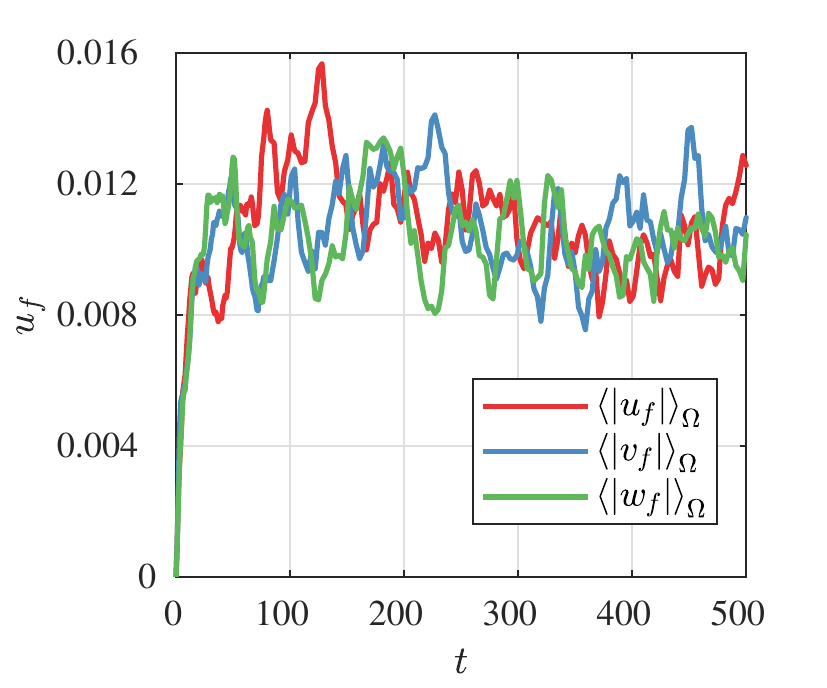}
    \label{fig:uMeanAbs_Tur1}
    \end{subfigure}
    \begin{subfigure}{0.5\textwidth}
    \centering
    \caption{}
    \includegraphics[width=0.9\textwidth]{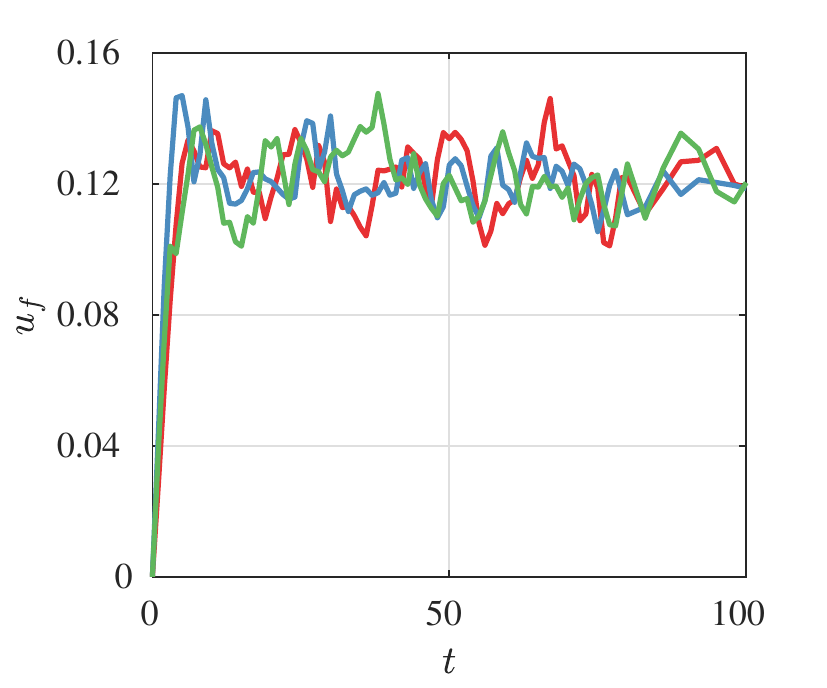}
    \label{fig:uMeanAbs_Tur8}
    \end{subfigure}
    \caption{Temporal evolution of box-averaged magnitude of the fluid velocity components: (a) Case Tur1; (b) Case Tur8. The flow is seen to be isotropic to a good approximation.}
    \label{fig:box-averaged absolute fluid velocity}
\end{figure}

Figure \ref{fig:box-averaged absolute fluid velocity} shows the temporal evolution of the domain-averaged magnitude of the velocity components $\langle |u_f| \rangle_{\Omega}$, $\langle |v_f| \rangle_{\Omega}$ and $\langle |w_f| \rangle_{\Omega}$. During the statistically stationary state the three components are seen to oscillate around similar average values for both Tur1 and Tur8, which indicates that the flow is isotropic to a good approximation.

\begin{figure}
    \begin{subfigure}{0.5\textwidth}
    \centering
    \caption{}
    \includegraphics[width=0.9\textwidth]{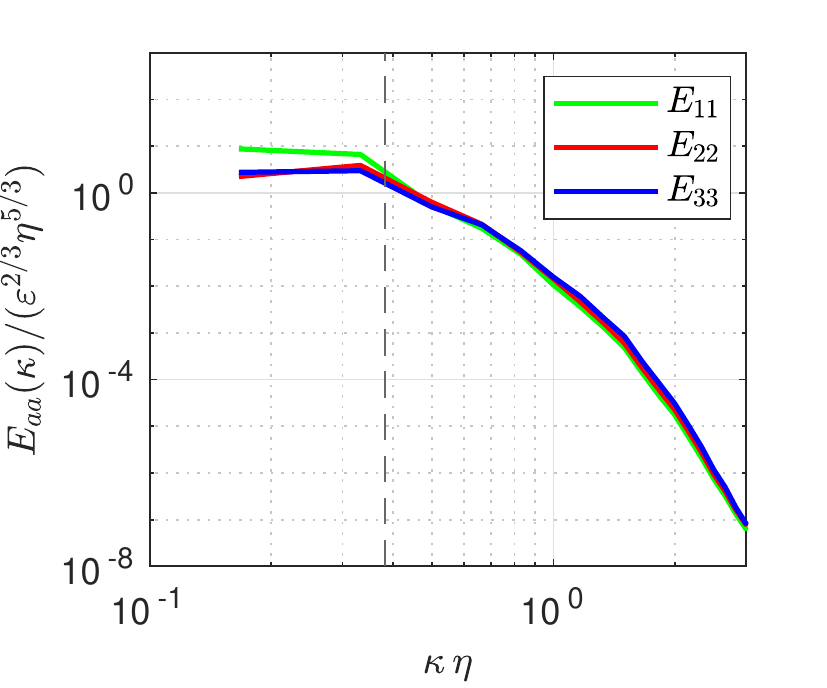}
    \label{fig:energyspectrum_Tur1}
    \end{subfigure}
    \begin{subfigure}{0.5\textwidth}
    \centering
    \caption{}
    \includegraphics[width=0.9\textwidth]{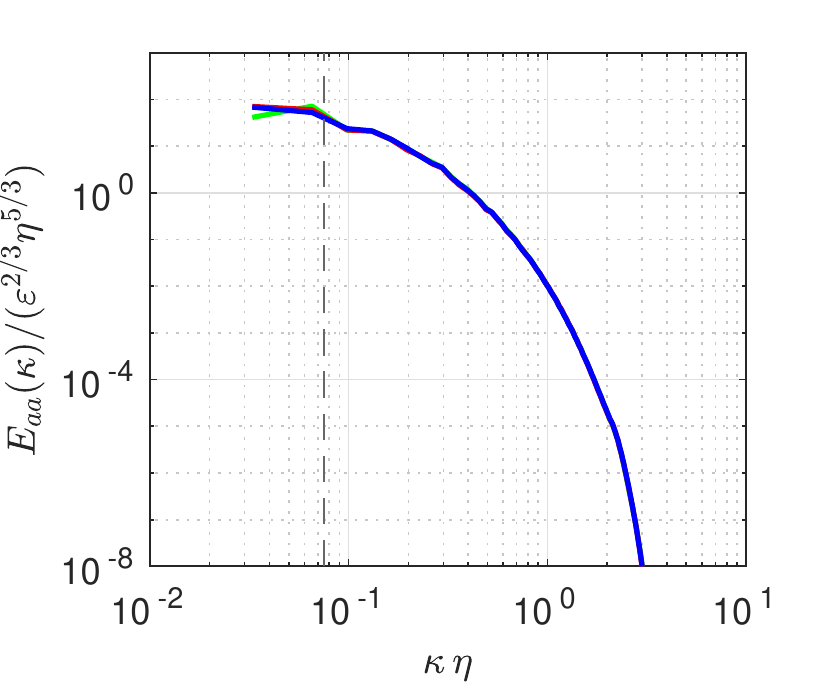}
    \label{fig:energyspectrum_Tur8}
    \end{subfigure}
    \caption{Time-averaged one-dimensional energy spectra. the vertical dashed lines indicate the respective cutoff wavenumber of the turbulence forcing scheme, $\kappa_f \, \eta = 2.3 (2\pi/L_x) \eta$. (a) Case Tur1; (b) Case Tur8. The spectra confirm that the statistically stationary flow fields are approximately isotropic.}
    \label{fig:time-averaged one-dimensional energy spectra}
\end{figure}

We define the instantaneous kinetic energy components in Fourier space, $E_{11}(\kappa)$, $E_{22}(\kappa)$ and $E_{33}(\kappa)$, as
\begin{subequations}
    \begin{eqnarray}
    \int_0^\infty E_{11}(\kappa)\, d \kappa = & \langle \frac{u_f \cdot u_f}{2} \rangle_{\Omega}   \ , \\
    \int_0^\infty E_{22}(\kappa)\, d \kappa = & \langle \frac{v_f \cdot v_f}{2} \rangle_{\Omega}   \ , \\
    \int_0^\infty E_{33}(\kappa)\, d \kappa = & \langle \frac{w_f \cdot w_f}{2} \rangle_{\Omega}   \ ,
    \end{eqnarray}
\end{subequations}
where $\kappa = |\boldsymbol \kappa|$ denotes the wavenumber. Figure \ref{fig:time-averaged one-dimensional energy spectra} shows the time-averaged one-dimensional energy spectra. Only the wavenumbers below the cutoff wavenumber ($\kappa_f$, shown as vertical dashed lines in Figure \ref{fig:time-averaged one-dimensional energy spectra}) are forced. The shapes of the energy spectra are in qualitative agreement with those obtained by \citet[p. 10]{chouippe2015forcing} for higher values of $Re_{\lambda} \approx 60$. We conclude that the present forcing scheme yields statistically steady flow fields that are approximately isotropic for the current range of $Re_{\lambda}$-values.

\section{Flocculation of cohesive particles} \label{sec:Flocculation of cohesive particles}

\subsection{One-way coupling} \label{Computational set-up}
Once the single-phase turbulence reaches the statistically stationary regime, $N = 10,000$ identical cohesive particles with diameter $D_p=0.008$ are randomly distributed throughout the domain, resulting in a particle volume fraction $\phi_p = 0.268\%$. Initially all particles are at rest and separated by a distance larger than the cohesive range $h_{co}$. \Rthree{To improve the statistics, we carry out repeated simulations for different random initial conditions, as the simulation results are statistically independent of the initial particle placement.} \Rone{The simulations to be discussed in the following are one-way coupled, so that the particles do not modify the background turbulence. \citet{bosse2006small} find that particle loading can modify the turbulence statistics even for volume fractions as low as $10^{-5}$, so that we expect two-way coupling effects to have an impact on the flocculation process even in moderately dilute flows. In addition, even for globally dilute flows the local volume fraction inside a floc will be $O(1)$, so that the one-way coupled assumption generally will not hold inside a floc. However, fully two-way coupled simulations for sufficiently many particles to obtain reliable statistical information, and for sufficiently long times to explore the balance between aggregation and breakup during the equilibrium stage, are not feasible on currently available supercomputers. Our assumption of one-way coupling hence limits the volume and mass fractions that we can reasonably consider.  On the other hand, the current simulations and their comparisons to experimental observations are useful in that they help address the question as to which aspects of flocculation are governed by one-way coupled dynamics, and which other aspects require fully two-way coupled dynamics. As we will see below, for the range of physical parameters listed in table \ref{tab:table1}, even one-way coupled simulations are able to reproduce several experimentally observed statistical features of flocculation dynamics.}

We adopt a multiscale time-stepping approach in which the fluid motion is calculated with a time step $\Delta t$ based on the criterion that the Courant-Friedrichs-Lewy number $\rm CFL \leq 0.5$. The particle motion, on the other hand, is evaluated with a much smaller time step $\Delta t_p = \Delta t / 15$. Since the computational approach maintains a contact duration of $T_c = 10 \Delta t = 150 \Delta t_p$ \citep{biegert2017collision}, each particle collision is effectively resolved by 150 substeps, at the price of a marginal increase in the computational cost. The dynamics of the primary particles are characterized by the Kolmogorov-scale Stokes number
\begin{equation} \label{eq:defination_St} 
    St = \frac{\rho_s}{18} \frac{D_p^2}{\eta^2} Re_{\eta} \ ,
\end{equation}
where the Kolmogorov Reynolds number $Re_{\eta} = \eta \, u_{rms} \, Re$. \Rtwo{Since the particle diameter $D_p$ is constant throughout the present investigation, $St$ depends on the density ratio $\rho_s$ and the fluid properties. A particle with a small Stokes number tends to follow the fluid motion, while the dynamics of a particle with a large Stokes number is dominated by its inertia, so that it tends to continue along its initial direction of motion.}

Table \ref{tab:flocculation cases} summarizes the physical parameters of the simulations that we conducted. \bernhard{Following our analysis from \S \ref{subsec:Non-dimensionalization} and the examples given in Appendix A of \cite{vowinckel2019settling}, these values correspond to primary silica particles with a grain size of fine to medium silt in ocean water.} In the following, we will investigate how the flocculation dynamics are influenced by the cohesive number $Co$, the Stokes number $St$ and the shear rate $G$. We remark that the density ratio $\rho_s$ and the size ratio $\eta/D_p$ are implicitly accounted for by $St$ and $G$.

\begin{table}
  \begin{center}
\def~{\hphantom{0}}
  \begin{tabular}{ccccccc}
    Case \ \ \ & Turbulent flow \ \ \ & $\eta/D_p$ \ \ \ & $\rho_s$ \ \ \ & $St$ \ \ \  & $Co$ \ \ \ & $G$   \\ 
    \\
    Flo1 \ \  & Tur1 & 3.30 & 2.65 & 0.02 & $\boldsymbol {4.0 \times 10^{-10}} $ & $0.29$   \\ 
    \\
    Flo2 \ \  & Tur1 & 3.30 & 2.65 & 0.02 & $\boldsymbol {6.0 \times 10^{-9}} $ & $0.29$   \\ 
    \\
    Flo3 \ \  & Tur1 & 3.30 & 2.65 & 0.02 & $\boldsymbol {3.0 \times 10^{-8}} $ & $0.29$   \\ 
    \\
    Flo4 \ \  & Tur1 & 3.30 & 2.65 & 0.02 & $\boldsymbol {6.0 \times 10^{-8}} $ & $0.29$   \\ 
    \\
    Flo5 \ \  & Tur1 & 3.30 & 2.65 & 0.02 & $\boldsymbol {1.2 \times 10^{-7}} $ & $0.29$   \\ 
    \\
    Flo6 \ \ & Tur2 & 2.28 & 2.65 & 0.06 & $1.2 \times 10^{-7}$ & $\boldsymbol {1.49}$   \\ 
    \\
    Flo7 \ \ & Tur3 & 2.25 & 2.65 & 0.06 & $1.2 \times 10^{-7}$ & $\boldsymbol {1.03}$   \\ 
    \\
    Flo8 \ \ & Tur4 & 2.24 & 2.65 & 0.06 & $1.2 \times 10^{-7}$ & $\boldsymbol {0.78}$   \\ 
    \\
    Flo9 \ \ & Tur5 & 2.25 & 2.65 & 0.06 & $1.2 \times 10^{-7}$ & $\boldsymbol {0.62}$   \\ 
    \\
    Flo10 \ \ & Tur6 & 1.85 & 2.65 & $\boldsymbol {0.10}$ & $1.2 \times 10^{-7}$ & $0.91$   \\ 
    \\
    Flo11 \ \ & Tur6 & 1.85 & 10 & $\boldsymbol {0.38}$ & $1.2 \times 10^{-7}$ & $0.91$   \\ 
    \\
    Flo12 \ \ & Tur6 & 1.85 & 20 & $\boldsymbol {0.77}$ & $1.2 \times 10^{-7}$ & $0.91$   \\ 
    \\
    Flo13 \ \ & Tur6 & 1.85 & 50 & $\boldsymbol {1.92}$ & $1.2 \times 10^{-7}$ & $0.91$   \\ 
    \\
    Flo14 \ \ & Tur7 & 1.08 & 2.65 & 0.38 & $1.2 \times 10^{-7}$ & $2.7$   \\ 
    \\
    Flo15 \ \ & Tur8 & 0.65 & 2.65 & 1.25 & $1.2 \times 10^{-7}$ & $7.4$   \\ 
    \\
  \end{tabular}
  \caption{Physical parameters of the flocculation simulations. We separately investigate the influence of the cohesive number $Co$ (based on Flo1-5), the shear rate $G$ (Flo6-9), and the Stokes number $St$ (Flo10-13). The effects of $\rho_s$ and $\eta/D_p$ are implicitly accounted for by $St$ and $G$.}
  \label{tab:flocculation cases}
  \end{center}
\end{table}

\subsection{Flocculation and equilibrium stages} \label{Two stages of flocculation: development and equilibrium}

\begin{figure}
    \begin{subfigure}{0.5\textwidth}
    \centering
    \caption{}
    \includegraphics[width=0.9\textwidth]{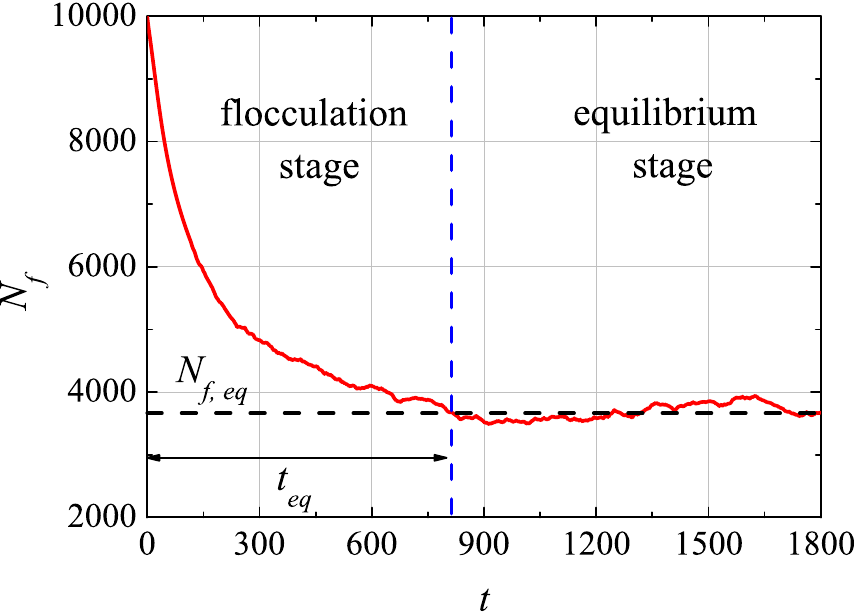}
    \label{fig:Nf_t}
    \end{subfigure}
    \begin{subfigure}{0.5\textwidth}
    \centering
    \caption{}
    \includegraphics[width=0.9\textwidth]{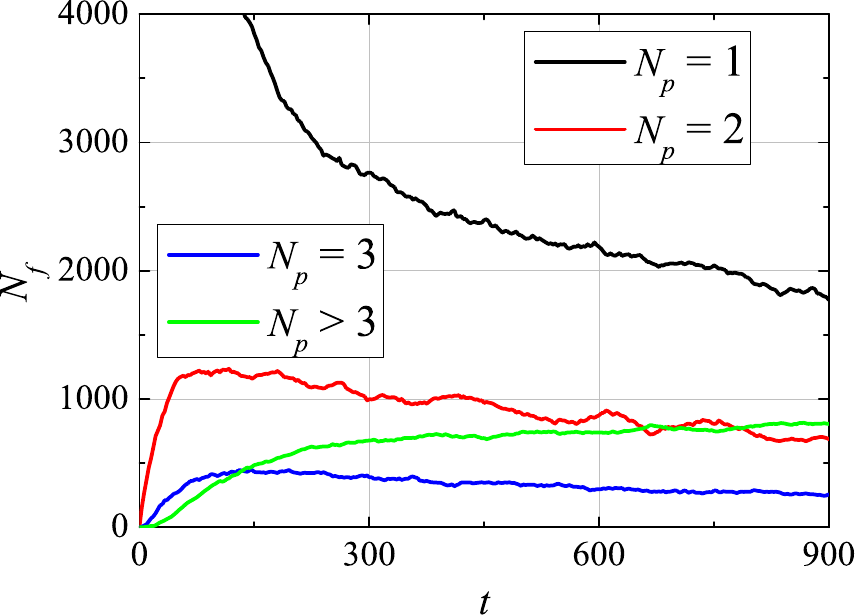}
    \label{fig:Nf_t_varying_Nplocal}
    \end{subfigure}
    \begin{subfigure}{1\textwidth}
    \centering
    \caption{}
    \includegraphics[width=0.45\textwidth]{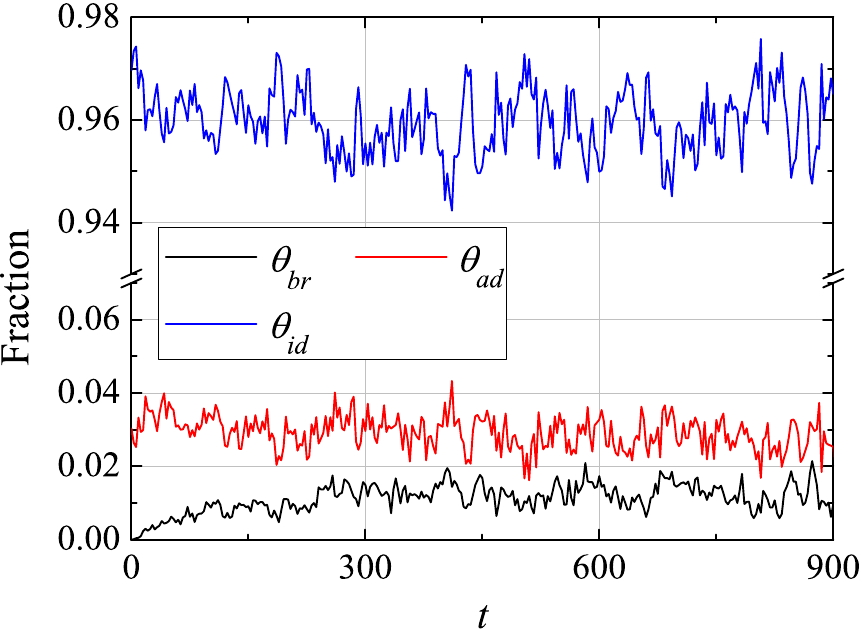}
    \label{fig:PDF_Nplocal}
    \end{subfigure}
    \caption{ (a) Temporal evolution of the number of flocs $N_f$. The vertical dashed line divides the simulation into the flocculation and equilibrium stages. (b) Number of flocs containing $N_p$ primary particles. The number of flocs with a single particle rapidly decreases from its initial value of $N_f = 10,000$. The numbers of flocs with two or three particles intially grow and subsequently decay, as increasingly many flocs with three or more particles form. (c) Temporal evolution of the fraction of flocs that maintain their identity ($\theta_{id}$), add primary particles ($\theta_{ad}$), or undergo breakage ($\theta_{br}$) over the time interval $\Delta T = 3$. All results are for case Flo9 with $Co = 1.2 \times 10^{-7}$, $St = 0.06$, $G = 0.62$, $\rho_s = 2.65$, $\eta/D_p = 2.25$.}
    \label{fig:two stages}
\end{figure}

When the surface distance between two particles is smaller than half the range of the cohesive force, $h_{co}/2$, we consider these particles to be part of the same floc. \bernhard{Hence, in terms of a physical force balance breakage occurs when the net force pulling the particles apart is sufficiently strong to overcome the maximum of the cohesive force holding the particles together.} An individual particle is considered to be the smallest possible floc. Figure \ref{fig:Nf_t} shows the evolution of the number of flocs $N_f(t)$ with time for the representative case Flo9, with $Co = 1.2 \times 10^{-7}$, $St = 0.06$, $G = 0.62$, $\rho_s = 2.65$ and $\eta/D_p = 2.25$. As a result of flocculation, $N_f$ decreases rapidly with time from its initial value of 10,000, before levelling off around a constant value $N_{f,eq}$ that reflects a stable equilibrium between aggregation and breakage. This tendency of $N_f$ is consistent with our previous observation of flocculation in steady cellular flow fields \citep{zhao2020efficient}. Consequently, we can identify two pronounced stages of the flow, viz. an initial flocculation stage and a subsequent equilibrium stage. We define the end of the flocculation stage, i.e., the onset of the equilibrium stage, as the time $t_{eq}$ when $N_f$ first equals $N_{f,eq}$. Figure \ref{fig:Nf_t_varying_Nplocal} shows separately the number of flocs with $N_{p} = 1, 2, 3$ and more than three primary particles. While the number of flocs with two or three particles initially grows quickly, they soon reach a peak and subsequently decline, as more flocs of larger sizes form. Toward the end of the flocculation stage, a stable equilibrium of the different floc sizes begins to emerge, although the distribution of flocs with different numbers of primary particles is still changing slowly.

In order to gain insight into the dynamics of floc growth and breakage, we keep track of the evolution of three different types of flocs over a suitably specified time interval $\Delta T$: a) those flocs that maintain their identity, i.e., they consist of the same primary particles at the start and the end of the time interval; b) those that add additional primary particles while keeping all of their original ones; and c) all others, i.e., all those who have undergone a breakage event during the time interval. We denote the fractions of these respective groups as $\theta_{id} = N_{f,id} / N_f$, $\theta_{ad} = N_{f,ad} / N_f$, and $\theta_{br} = N_{f,br} / N_f$. It follows that
\begin{equation} \label{eq:PDF flocs 1}
    \theta_{id} + \theta_{ad} + \theta_{br} = 1 \ .
\end{equation}
We found that a value of $\Delta T = 3$ is suitable for obtaining insight into the dynamics of the flocculation process, as it allows most of the flocs to maintain their identity during the time interval, while smaller but still significant numbers undergo primary particle addition or breakage. Figure \ref{fig:PDF_Nplocal} shows the evolution of $\theta_{id}$, $\theta_{ad}$ and $\theta_{br}$ for case Flo9. After an initial transient stage, all three fractions reach statistically steady states. Interestingly, even during the equilibrium stage when $N_f \approx const.$, we observe that $\theta_{ad} \ne \theta_{br}$. This reflects events such as when one floc breaks into three smaller parts, two of which then merge with other flocs. Here the total number of flocs remains unchanged at three, in spite of only one break-up but two particle addition events.

\subsection{Evolution of floc size and shape} \label{Temporal evolution of the size of flocs}

While the number of primary particles in a floc, $N_{p}$, provides a rough measure of its size, flocs with identical values of $N_p$ can have very different shapes. In order to capture this effect, we define the characteristic diameter $D_f$ of the floc, also known as the Feret diameter, as
\begin{equation}\label{eq:defination_Df}
    D_f = 2 {\rm max}(\| \boldsymbol x_{p,i} - \boldsymbol x_c \|) + D_p, \ \ \ 1 \leqslant i \leqslant N_{p} \ ,
\end{equation}
as well as its gyration diameter $D_g$ \citep{chen2019exponential},
\begin{equation}
    D_g = \left \{
    \begin{array}{lll}
    2\sqrt{\frac{1}{N_{p}} \sum_{i=1}^{N_{p}} \| \boldsymbol x_{p,i} - \boldsymbol x_c \|^2}, & N_{p} > 2 \ , \\
    \\
    \sqrt{1.6}D_p, & N_{p} = 2 \ , \\ 
    \\
    D_p, & N_{p} = 1 \ . \\
    \end{array}\right.
\end{equation}
Here $\boldsymbol x_{p,i}$ denotes the position of the center of primary particle $i$, and the floc's center of mass is evaluated as $\boldsymbol x_c = \sum_{i=1}^{N_{p}} \boldsymbol x_{p,i} / N_{p}$. While the characteristic diameter $D_f$ more closely represents the true spatial extent of the floc, the gyration diameter $D_g$ also accounts for the irregularity of the floc shape. 

Following \citet{Khelifa2006II, Khelifa2006I}, we then calculate the fractal dimension $n_f$ of the floc
\begin{equation} \label{eq:defination_nf} 
    n_f = \frac{\log N_{p}}{\log \frac{D_f}{D_p}} \ ,
\end{equation}
as a measure of its compactness. A dense, nearly spherical floc has $n_f \approx 3$, while for a linear floc $n_f \approx 1$. When $N_{p} = 1$ and $D_f/D_p = 1$, the above definition of the fractal dimension does not yield a finite value, and we set $n_f = 1$. In this way, the definition of the fractal dimension is continuous between $N_p=1$ and $N_p=2$. It is important to note that this differs from previous studies, which usually set $n_f = 3$ for this case \citep{Khelifa2006II, Khelifa2006I, maggi2007effect, son2009effect}.

\begin{figure}
    \centering
    \includegraphics[width=0.75\textwidth]{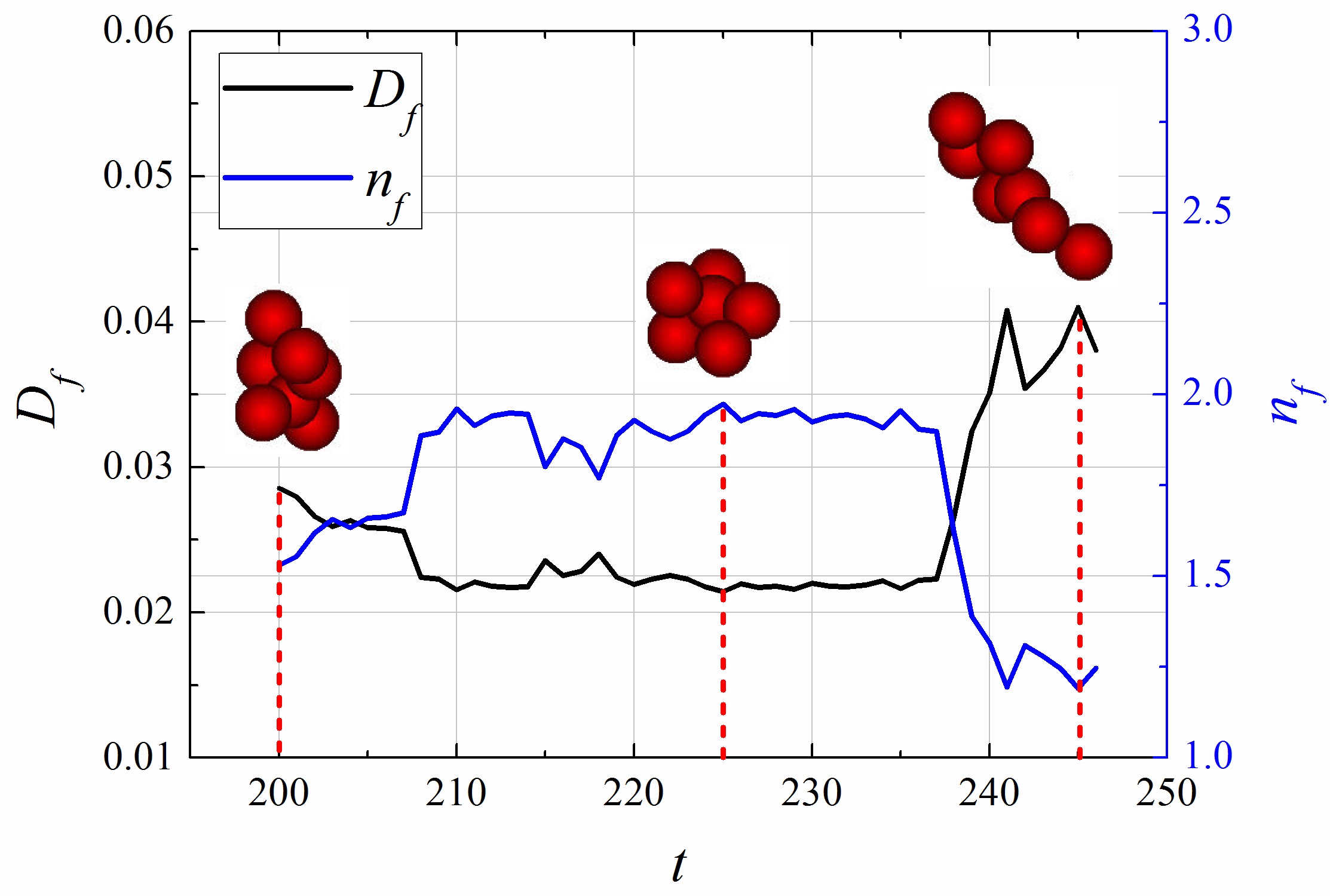}
    \caption{Temporal evolution of the characteristic diameter $D_f$ and the fractal dimension $n_f$ of a typical floc that maintains its identity over the time interval considered. Three instants are marked by vertical dashed lines, and the corresponding floc shapes are shown. In response to the fluid forces acting on it, the floc first changes from a slightly elongated to a more compact shape, and subsequently to a more strongly elongated one. The floc with seven primary particles is taken from case Flo10 with governing parameters $Co = 1.2 \times 10^{-7}$, $St = 0.1$, $G = 0.91$.}
    \label{fig:typical_maintain_floc_evolution}
\end{figure}

For a typical floc with $N_p = 7$ that maintains its identity, figure \ref{fig:typical_maintain_floc_evolution} shows the evolution of $D_f$ and $n_f$ over time. During the time interval $200 \leqslant t \leqslant 210$, hydrodynamic forces deform the floc so that it becomes more compact, which reduces $D_f$ and increases $n_f$. Later on, near $t \approx 240$, the floc is being stretched, which modifies $D_f$ and $n_f$ in the opposite directions.

\begin{figure}
    \begin{subfigure}{0.5\textwidth}
    \centering
    \caption{}
    \includegraphics[width=0.9\textwidth]{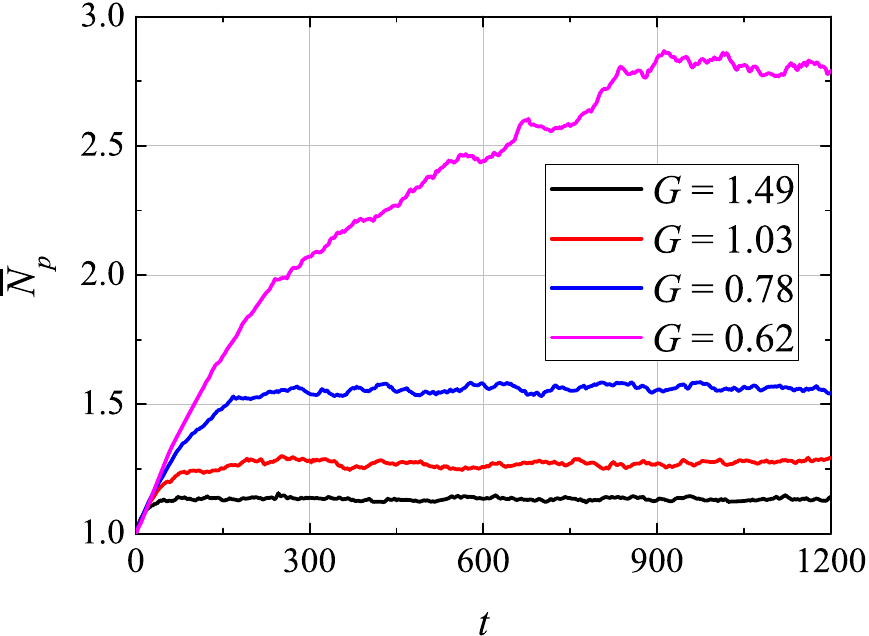}
    \label{fig:Nplocal_t_varying_G}
    \end{subfigure}
    \begin{subfigure}{0.5\textwidth}
    \centering
    \caption{}
    \includegraphics[width=0.9\textwidth]{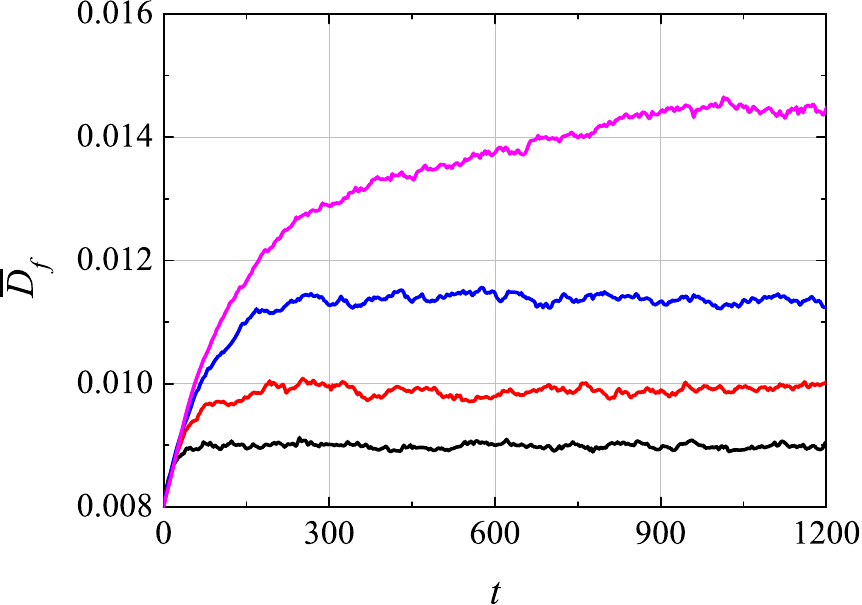}
    \label{fig:Df_t_varying_G}
    \end{subfigure}
    \begin{subfigure}{0.5\textwidth}
    \centering
    \caption{}
    \includegraphics[width=0.9\textwidth]{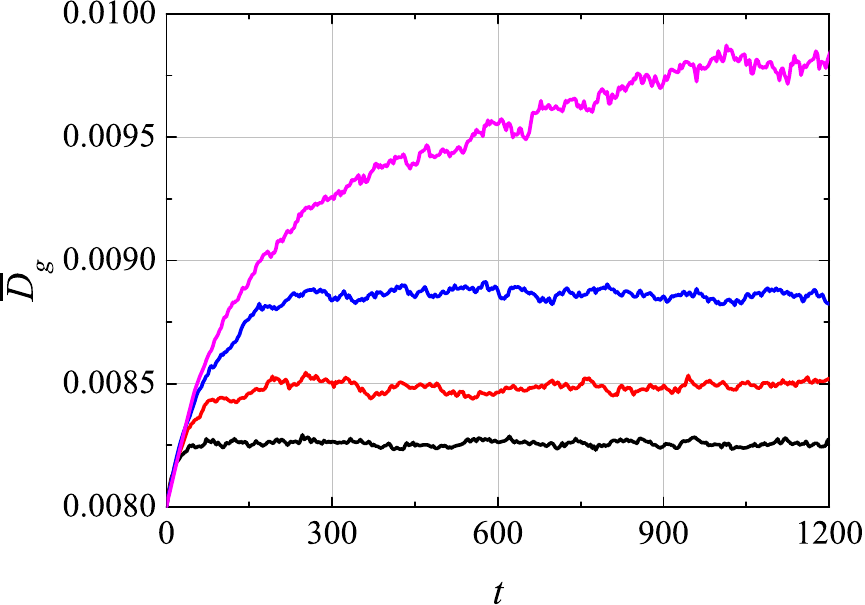}
    \label{fig:Dg_t_varying_G}
    \end{subfigure}
    \begin{subfigure}{0.5\textwidth}
    \centering
    \caption{}
    \includegraphics[width=0.9\textwidth]{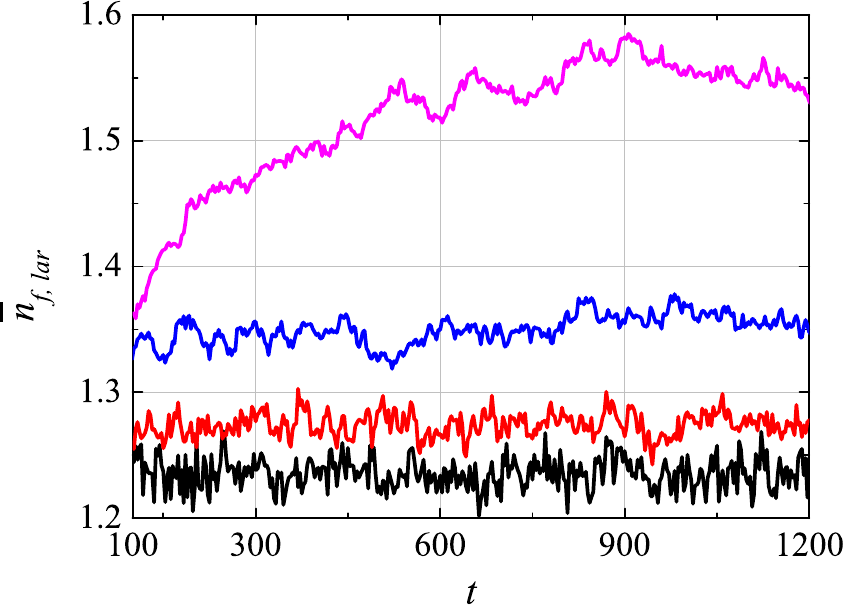}
    \label{fig:nnf_more_t_varying_G}
    \end{subfigure}
    \caption{Temporal evolution of various floc size measures for different turbulent shear rates $G$, with $Co = 1.2 \times 10^{-7}$, $St = 0.06$, $\rho_s = 2.65$, and $2.24 \leqslant \eta/D_p \leqslant 2.28$ (cases Flo6-9). (a) Average number of primary particles per floc $\overline N_p$; (b) Average characteristic floc diameter $\overline D_f$; (c) Average floc gyration diameter $\overline D_g$; (d) Average fractal dimension $\overline n_{f,lar}$ of flocs with three or more primary particles. Larger turbulent shear results in smaller flocs, with fewer primary particles and more elongated shapes.}
    \label{fig:temporal evolution of floc size G}
\end{figure}

Figure \ref{fig:temporal evolution of floc size G} shows the evolution with time of the various floc size measures, for cases Flo6-9 in table \ref{tab:flocculation cases} with different turbulent shear rates $G$. The other parameters are kept approximately constant at $Co = 1.2 \times 10^{-7}$, $St = 0.06$, $\rho_s = 2.65$, and $2.24 \leqslant \eta/D_p \leqslant 2.28$. As can be seen from figure \ref{fig:Nplocal_t_varying_G}, a smaller shear rate results in a longer transient phase before the average number of primary particles per floc $\overline N_{p} = N / N_f$ reaches an equilibrium. A smaller value of $G$ furthermore gives rise to an equilibrium stage characterized by fewer flocs with more primary particles, since the weaker hydrodynamic stresses cannot break up the flocs as easily. Figures \ref{fig:Df_t_varying_G} and \ref{fig:Dg_t_varying_G} indicate that both the average characteristic diameter $\overline D_f$ and the average gyration diameter $\overline D_g$ increase for smaller $G$. This is consistent with previous observations by other authors in both laboratory experiments \citep{he2012characteristic, guerin2017dynamics} and river estuaries \citep{manning2002use, manning2010review}. Both $\overline D_g$ and $\overline D_f$ remain smaller than the Kolmogorov length scale $0.0179 \leqslant \eta \leqslant 0.0183$ for all cases. Since flocs with one or two primary particles have a constant fractal dimension $n_f = 1$, we evaluate the average fractal dimension $\overline n_{f,lar}$ from only those flocs with three or more particles. Figure \ref{fig:nnf_more_t_varying_G} shows that $\overline n_{f,lar}$ increases for smaller shear rates, which demonstrates that for weaker turbulence the floc shape tends to be more compact. This finding is consistent with experimental observations by \citet{he2012characteristic}, whereas previous numerical work by \citet{chen2019exponential} reports a constant value $\overline n_{f,lar} = 1.64$.

\begin{figure}
    \begin{subfigure}{0.5\textwidth}
    \centering
    \caption{}
    \includegraphics[width=0.9\textwidth]{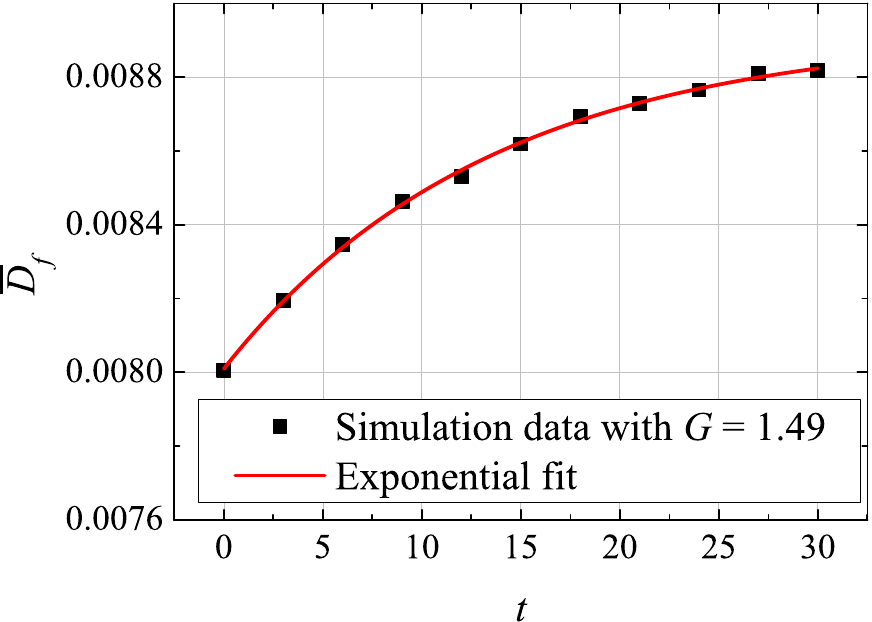}
    \label{fig:ExpDecay_Fit}
    \end{subfigure}
    \begin{subfigure}{0.5\textwidth}
    \centering
    \caption{}
    \includegraphics[width=0.9\textwidth]{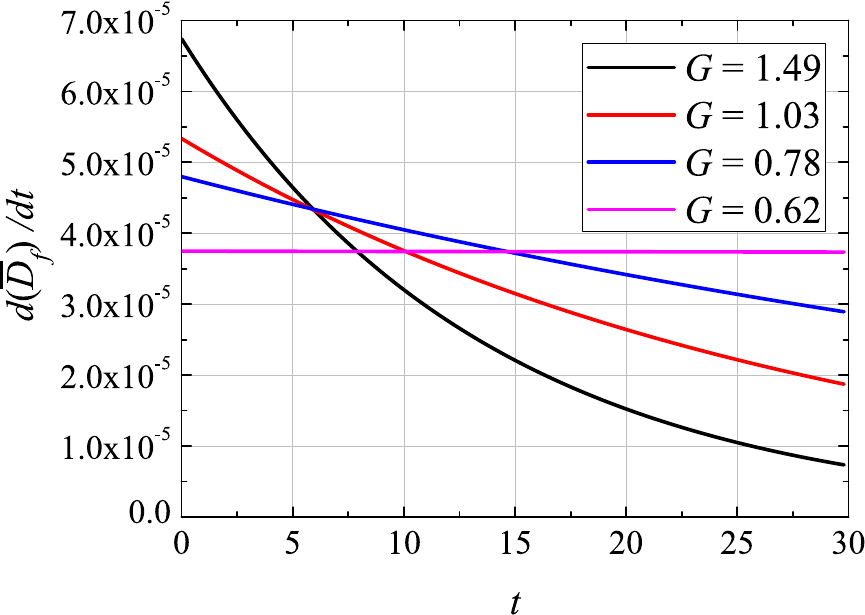}
    \label{fig:Df_speed_t_varying_G}
    \end{subfigure}
    \caption{Early-stage flocculation rate for different turbulent shear rates $G$, with $Co = 1.2 \times 10^{-7}$, $St = 0.06$, $\rho_s = 2.65$, and $\eta/D_p \approx 2.26$ (cases Flo6-9). (a) The early-stage simulation results for $\overline D_f(t)$ can be accurately fitted by an exponential relation, as shown for the representative case Flo6 with $G = 1.49$; (b) The flocculation rate $d(\overline D_f)/dt$ obtained from the exponential fits of $\overline D_f(t)$. Initially flocs grow fastest in strong turbulence. Subsequently their growth rate decays, as the equilibrium stage is reached more rapidly for strong turbulence.}
    \label{fig:temporal evolution of floc rate G}
\end{figure}

Figure \ref{fig:temporal evolution of floc rate G} discusses the floc growth during the very early flow stages, as a function of the turbulent shear rate $G$. As seen in Figure \ref{fig:ExpDecay_Fit}, the evolution of $\overline D_f(t)$ can be closely approximated by an exponential function of the form
\begin{equation} \label{eq:fitting_Df} 
    \overline D_f = b_1 (e^{b_2 t} - 1) + D_p \ ,
\end{equation}
where $b_1$ and $b_2$ represent fitting coefficients. Based on corresponding fits for different values of $G$, Figure \ref{fig:Df_speed_t_varying_G} displays the time-dependent floc growth rate $d\overline D_f /dt$ for different $G$. Consistent with the experimental observations by \citet{he2012characteristic}, we find stronger shear to cause more rapid flocculation for $t < 5$. After this early stage the trends reverse, which reflects the fact that the equilibrium stage is reached faster for stronger turbulence. This agrees with the experimental findings by \citet{braithwaite2012controls}, who also reported the equilibrium stage to emerge more quickly for stronger turbulence, due to more frequent floc collisions. We remark that the evolution of $\overline N_{p}$ (not shown) exhibits corresponding trends.

\begin{figure}
    \begin{subfigure}{0.5\textwidth}
    \centering
    \caption{}
    \includegraphics[width=0.9\textwidth]{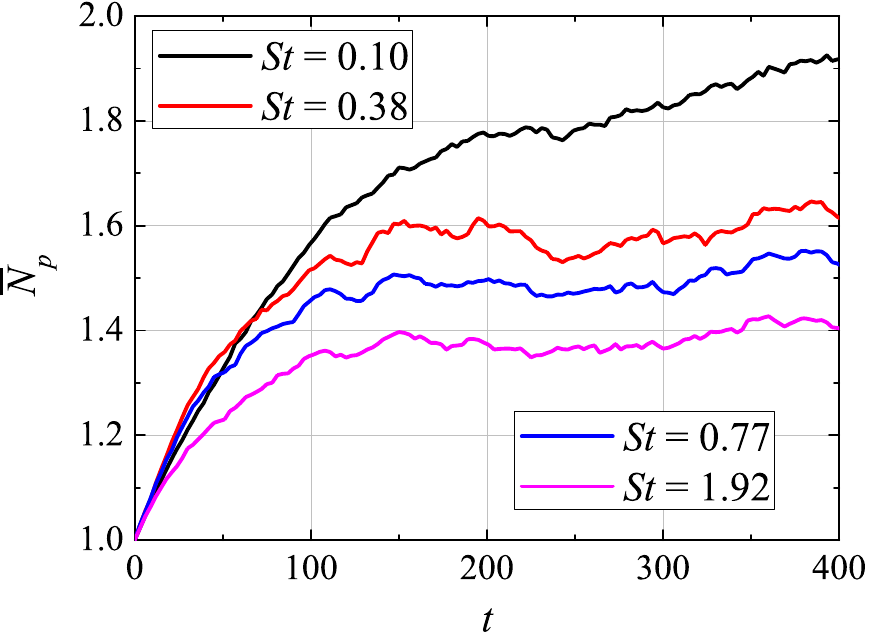}
    \label{fig:Nplocal_t_varying_St}
    \end{subfigure}
    \begin{subfigure}{0.5\textwidth}
    \centering
    \caption{}
    \includegraphics[width=0.9\textwidth]{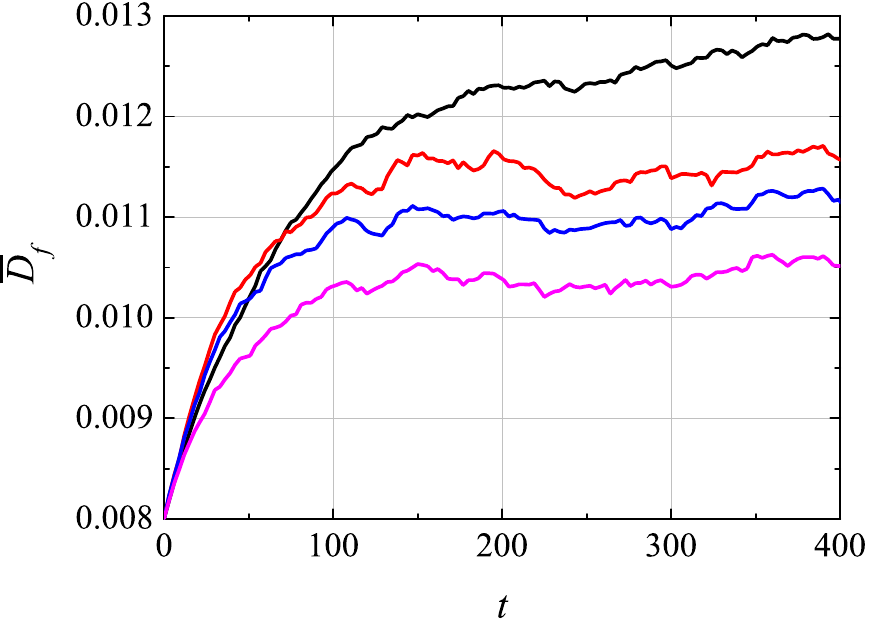}
    \label{fig:Df_t_varying_St}
    \end{subfigure}
    \begin{subfigure}{0.5\textwidth}
    \centering
    \caption{}
    \includegraphics[width=0.9\textwidth]{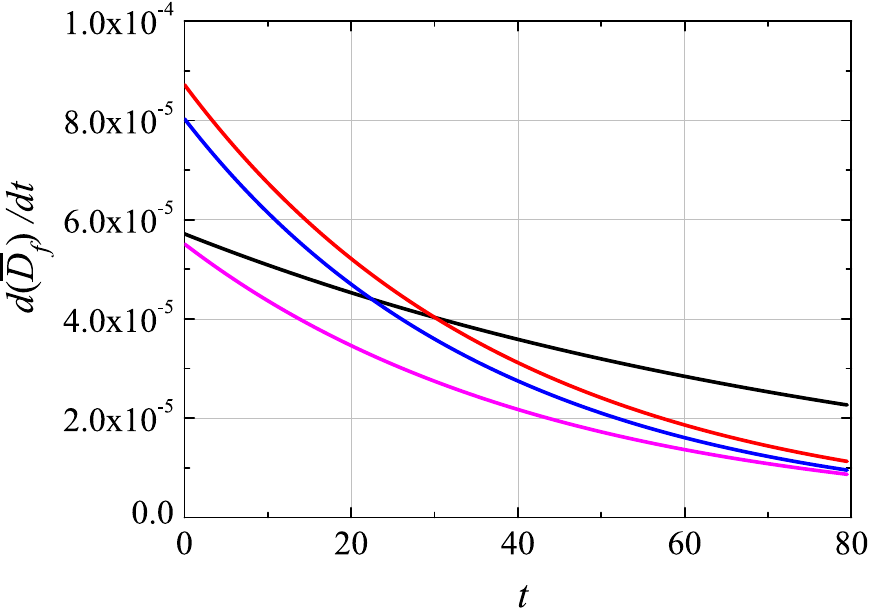}
    \label{fig:Df_speed_t_varying_St}
    \end{subfigure}
    \begin{subfigure}{0.5\textwidth}
    \centering
    \caption{}
    \includegraphics[width=0.9\textwidth]{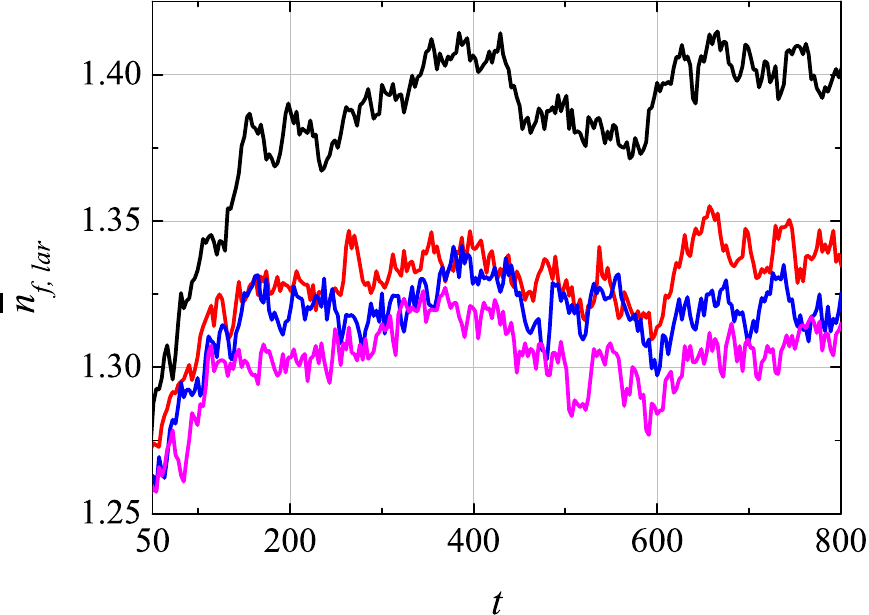}
    \label{fig:nnf_more_t_varying_St}
    \end{subfigure}
    \caption{Temporal evolution of various floc size measures, for different Stokes number values $St$, with $Co = 1.2 \times 10^{-7}$, $G = 0.91$ and $\eta/D_p = 1.85$ (cases Flo10-13). (a) Average number of primary particles per floc $\overline N_p$; (b) Average characteristic floc diameter $\overline D_f$; (c) Early-stage flocculation rate $d(\overline D_f)/dt$ obtained from exponential fits of $\overline D_f(t)$; (d) Average fractal dimension $\overline n_{f,lar}$ of flocs with three or more primary particles. During the equilibrium stage, the number of primary particles per floc, the characteristic floc diameter, and the fractal dimension all increase for smaller Stokes numbers. Initially, flocs with $St \approx O(1)$ exhibit the fastest growth.}
    \label{fig:temporal evolution of floc St}
\end{figure}

Figure \ref{fig:temporal evolution of floc St} presents corresponding floc size results for different Stokes numbers, obtained from cases Flo10-13 in table \ref{tab:flocculation cases}. These simulations all employ the same turbulent flow Tur6, so that they have constant parameter values $Co = 1.2 \times 10^{-7}$, $G = 0.91$ and $\eta/D_p = 1.85$. $St$ is varied by changing the density ratio $\rho_s$. Figures \ref{fig:Nplocal_t_varying_St} and \ref{fig:Df_t_varying_St} indicate that the equilibrium values of both $\overline N_{p}$ and $\overline D_f$ increase for smaller $St$. This reflects the fact that cohesive forces become more dominant for smaller $St$, due to the lower drag force and the shorter particle response time. By again employing exponential fits for the early stages, we obtain the floc growth rate $d\overline D_f / dt$ for different $St$-values, as shown in Figure \ref{fig:Df_speed_t_varying_St}. Initially flocs with intermediate Stokes numbers of \textit{O}(1) are seen to grow most rapidly, consistent with our earlier findings for two-dimensional cellular flows \citep{zhao2020efficient}. This trend changes for $t > 20$, due to the later onset of the equilibrium stage for small Stokes numbers. The time evolution of the average fractal dimension $\overline n_{f,lar}$ of flocs with three or more primary particles is shown in Figure \ref{fig:nnf_more_t_varying_St}. It demonstrates that smaller Stokes numbers result in more compact flocs.

\begin{figure}
    \begin{subfigure}{0.5\textwidth}
    \centering
    \caption{}
    \includegraphics[width=0.9\textwidth]{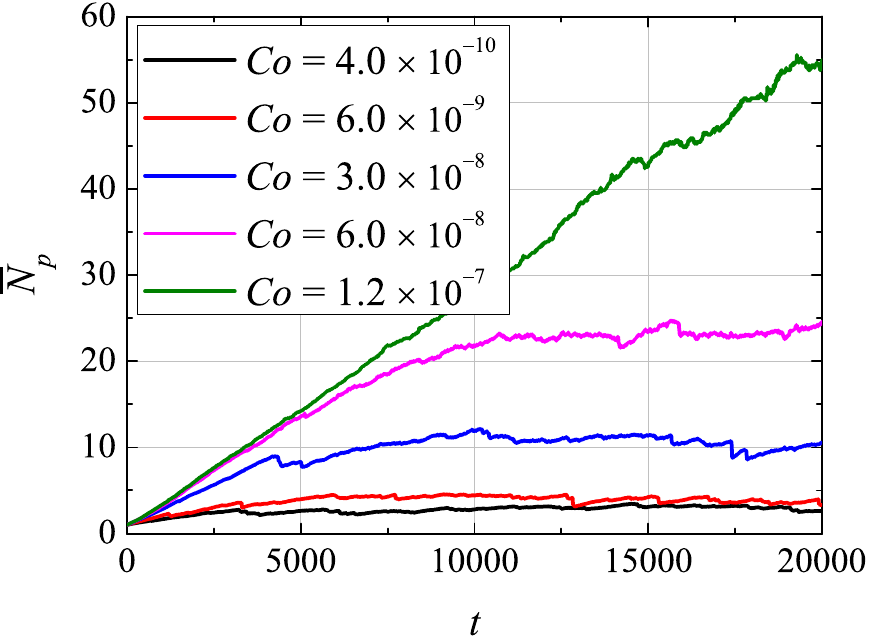}
    \label{fig:Nplocal_t_varying_Co}
    \end{subfigure}
    \begin{subfigure}{0.5\textwidth}
    \centering
    \caption{}
    \includegraphics[width=0.9\textwidth]{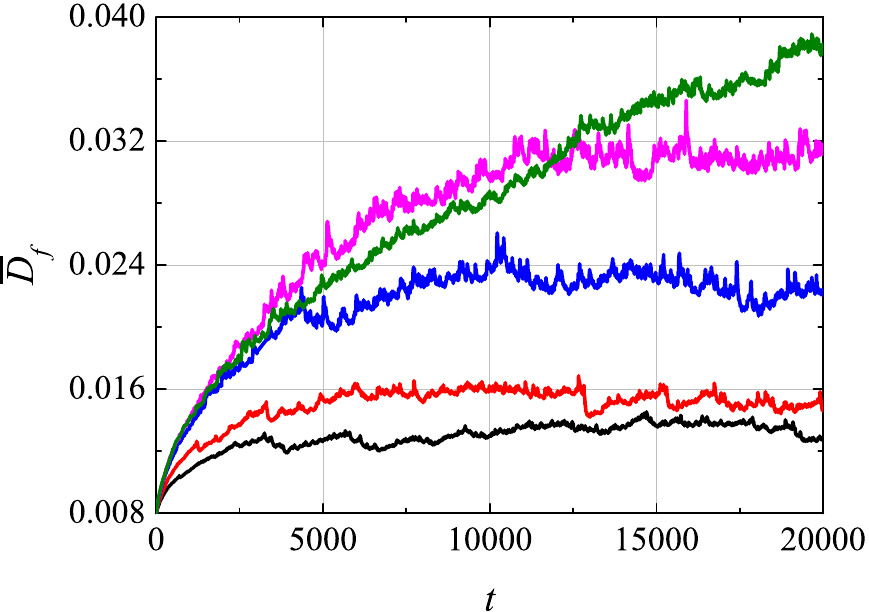}
    \label{fig:Df_t_varying_Co}
    \end{subfigure}
    \begin{subfigure}{0.5\textwidth}
    \centering
    \caption{}
    \includegraphics[width=0.9\textwidth]{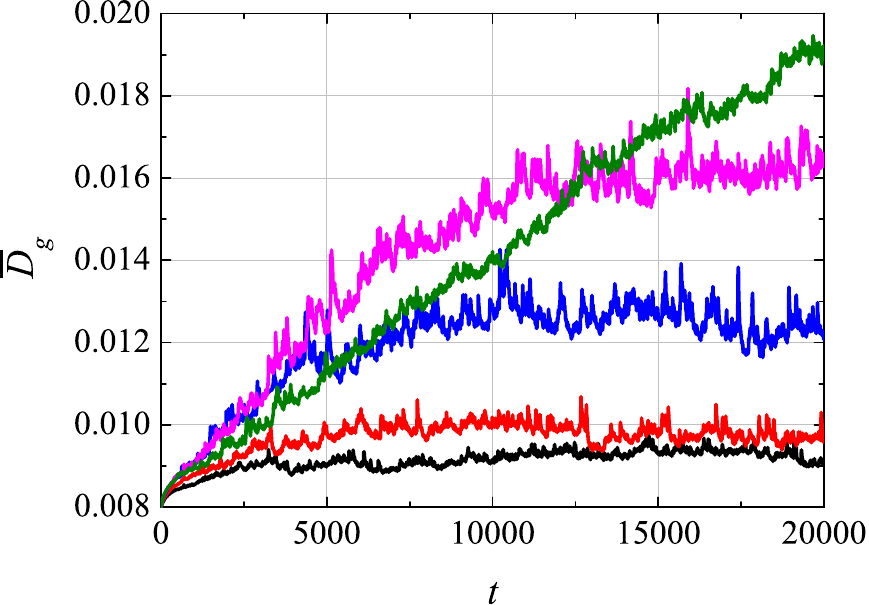}
    \label{fig:Dg_t_varying_Co}
    \end{subfigure}
    \begin{subfigure}{0.5\textwidth}
    \centering
    \caption{}
    \includegraphics[width=0.9\textwidth]{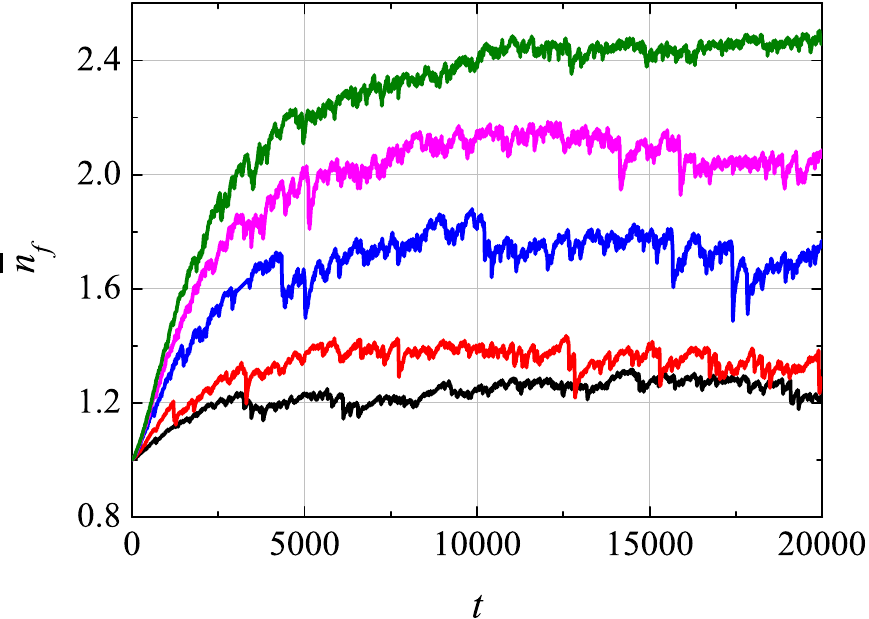}
    \label{fig:nnf_t_varying_Co}
    \end{subfigure}
    \caption{Temporal evolution of various floc size measures for different values of the cohesive number $Co$, with $St = 0.02$, $G = 0.29$, $\rho_s = 2.65$ and $\eta/D_p = 3.30$ (cases Flo1-5). (a) Average number of primary particles per floc $\overline N_p$; (b) Average characteristic floc diameter $\overline D_f$; (c) Average floc gyration diameter $\overline D_g$; (d) Average fractal dimension of flocs $\overline n_f$. Note that case Flo5 with $Co = 1.2 \times 10^{-7}$ has not yet reached the equilibrium stage by the end of the simulation. For higher $Co$-values, the equilibrium stage is characterized by larger flocs with more primary particles. During the transient stages, however, intermediate $Co$-values can give rise to flocs that are more elongated and hence larger than those at higher $Co$-values, in spite of having fewer primary particles.}
    \label{fig:temporal evolution of floc Co}
\end{figure}

Figure \ref{fig:temporal evolution of floc Co} analyzes the influence of the cohesive number $Co$ by comparing cases Flo1-5 in table \ref{tab:flocculation cases}. The other parameters are held constant at $St = 0.02$, $G = 0.29$ $\rho_s = 2.65$ and $\eta/D_p = 3.30$. We note that due to the small values of $St$ and $G$, the emergence of an equilibrium stage takes longer in these simulations. In fact, for case Flo5 with $Co = 1.2 \times 10^{-7}$, an equilibrium had not yet formed by $t=20,000$, when the simulation terminated. Nevertheless, the simulations demonstrate the tendency of higher $Co$ to result in larger values of $\overline N_{p}$ during all phases of the flow, cf. Figure \ref{fig:Nplocal_t_varying_Co}. Interestingly, however, we observe that during the transient flow stages the flocs for $Co = 6 \times 10^{-8}$ have larger average diameters $\overline D_f$ and $\overline D_g$ than those for $Co = 1.2 \times 10^{-7}$, even though they contain fewer primary particles, cf. Figures \ref{fig:Df_t_varying_Co} and \ref{fig:Dg_t_varying_Co}. The explanation for this finding is given by Figure \ref{fig:nnf_t_varying_Co}, which indicates that for $Co = 1.2 \times 10^{-7}$ the flocs have a higher average fractal dimension $\overline n_f$ and are more compact than those for $Co = 6 \times 10^{-8}$, which can be deformed more easily by turbulent stresses.

In summary, as a general trend we observe that during the equilibrium stages weaker turbulence, lower Stokes numbers and higher cohesive numbers result in larger and more compact flocs.

\subsection{Floc size distribution during the equilibrium stage} \label{Floc size distribution}

\begin{figure}
    \begin{subfigure}{0.5\textwidth}
    \centering
    \caption{}
    \includegraphics[width=0.9\textwidth]{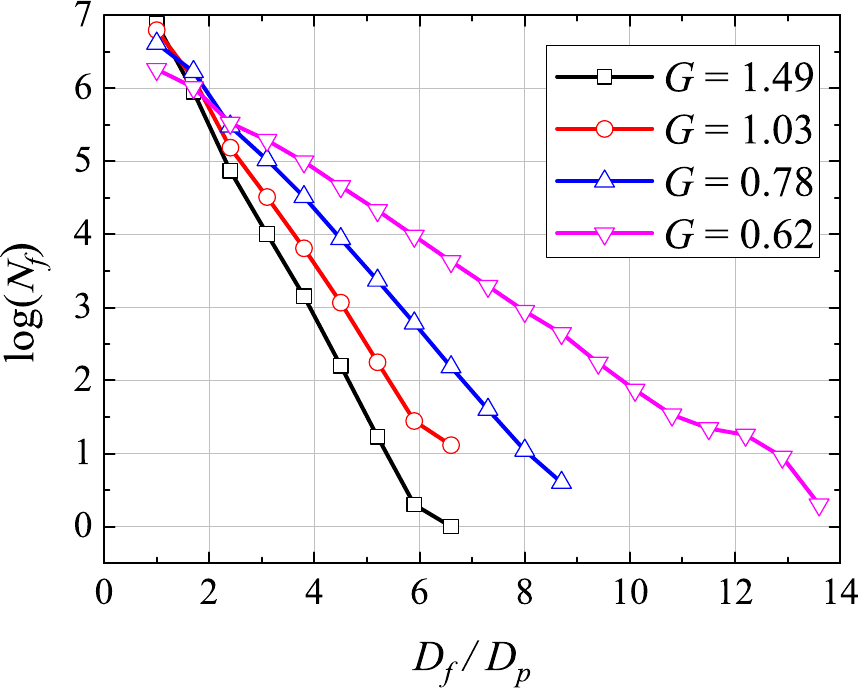}
    \label{fig:floc_size_distribution_varying_G_log}
    \end{subfigure}
    \begin{subfigure}{0.5\textwidth}
    \centering
    \caption{}
    \includegraphics[width=0.9\textwidth]{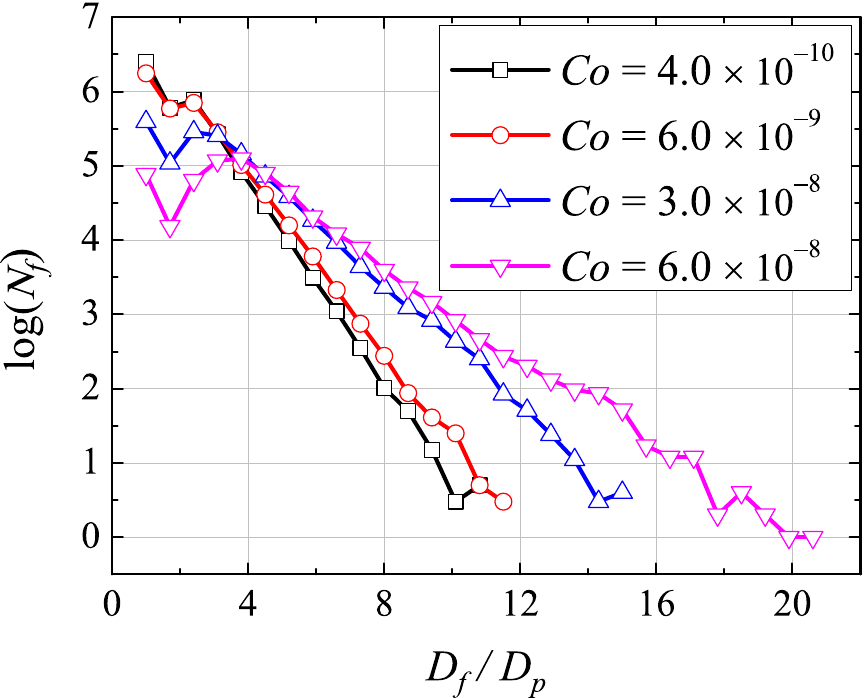}
    \label{fig:floc_size_distribution_varying_Co_log}
    \end{subfigure}
    \caption{Floc size distribution during the equilibrium stage, obtained by sorting all flocs into bins of constant width $\Delta(D_f/D_p) = 0.7$. (a) Results for different shear rates $G$, with $Co = 1.2 \times 10^{-7}$ and $St = 0.06$, during the time interval $1,000 \leqslant t \leqslant 4,000$ (cases Flo6-9); (b) Results for different cohesive numbers $Co$, with $St = 0.02$ and $G = 0.29$, for the time interval $15,000 \leqslant t \leqslant 19,000$ (cases Flo1-4).}
    \label{fig:floc_size_distribution}
\end{figure}

In order to discuss the floc size distribution during the equilibrium stage, we sort the flocs into bins of width $\Delta(D_f/D_p) = 0.7$. Figure \ref{fig:floc_size_distribution_varying_G_log} shows that for all values of the turbulent shear $G$ the size distribution peaks at the smallest flocs and then decreases exponentially with the floc size. The decay rate is largest for the strongest turbulence, confirming our earlier observation that strong turbulence breaks up large flocs and reduces the average floc size, cf. Figure \ref{fig:temporal evolution of floc size G}. This finding is consistent with the experimental observations by \citet{braithwaite2012controls} in an energetic tidal channel. Corresponding results for different $St$-values display a similar trend (not shown).

Figure \ref{fig:floc_size_distribution_varying_Co_log} shows the size distributions for different values of the cohesive number. For larger values of $Co$, we find that the peak of the distribution decreases and shifts to larger flocs, while the exponential decay rate with increasing floc size is reduced.

\subsection{\Rtwo{Change in floc microstructure}} \label{Deformation of flocs}

\begin{figure}
    \begin{subfigure}{0.5\textwidth}
    \centering
    \caption{}
    \includegraphics[width=0.9\textwidth]{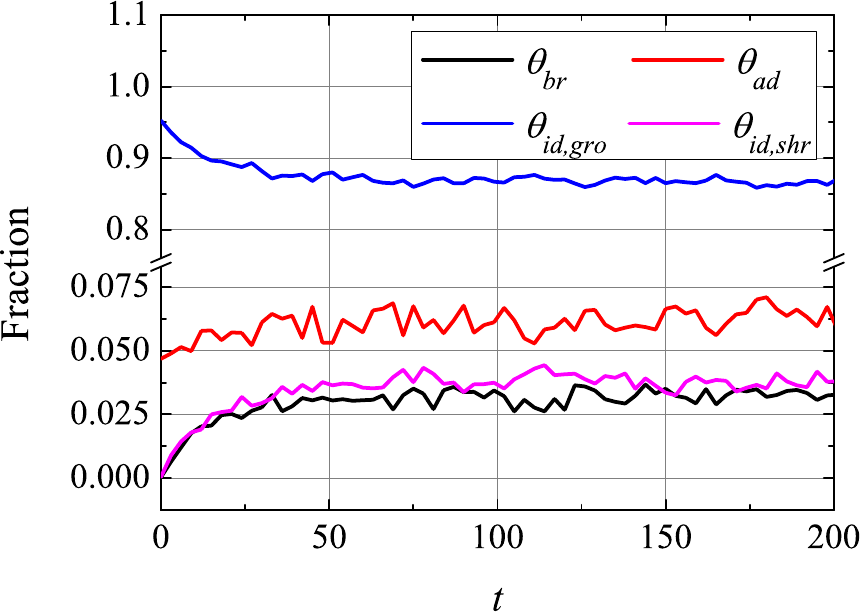}
    \label{fig:PDF_Re2e3_Ds5.5e4}
    \end{subfigure}
    \begin{subfigure}{0.5\textwidth}
    \centering
    \caption{}
    \includegraphics[width=0.9\textwidth]{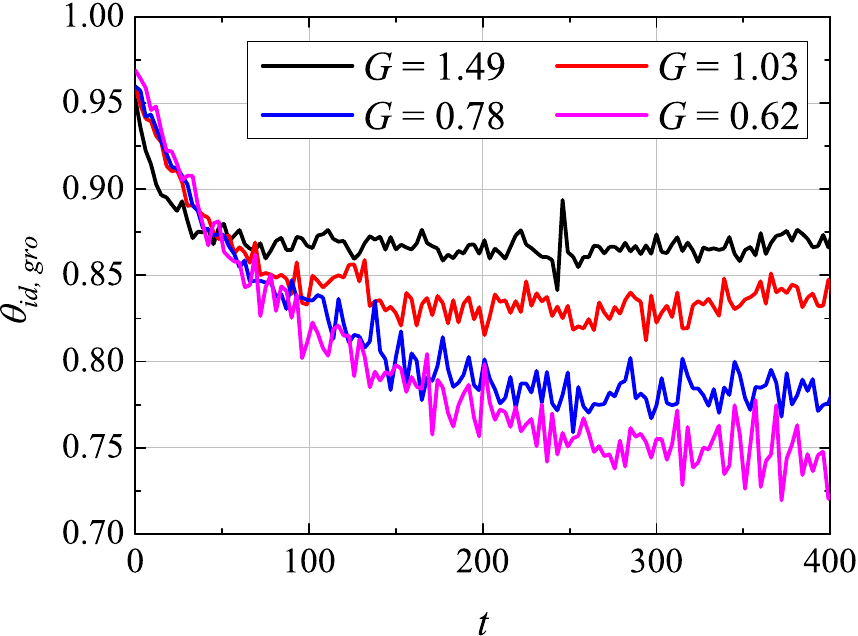}
    \label{fig:theta_mai_gro_varing_G}
    \end{subfigure}
    \begin{subfigure}{1\textwidth}
    \centering
    \caption{}
    \includegraphics[width=0.45\textwidth]{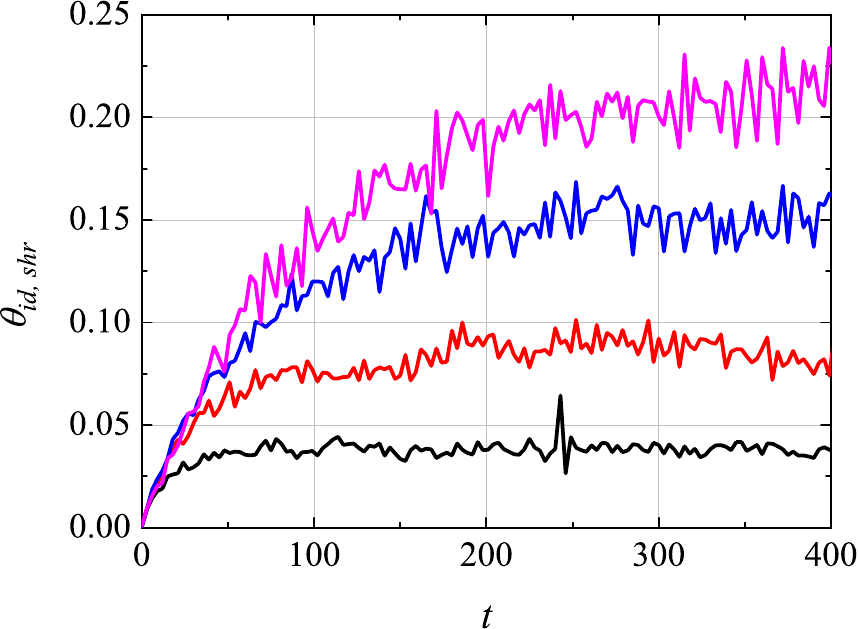}
    \label{fig:theta_mai_shr_varing_G}
    \end{subfigure}
    \caption{Evolution of the floc number fractions displaying different behaviors. (a) Of those flocs that maintain their identity during $\Delta T$, many more are being stretched than shrink, resulting in $\theta_{id,gro} \gg \theta_{id,shr}$ (case Flo6 with $G = 1.49$); (b) The fraction $\theta_{id,gro}$ that is being stretched increases for more intense turbulence; (c) The fraction $\theta_{id,shr}$ that shrinks decreases for stronger turbulence. For (b) and (c) the color coding of the curves is identical, and the other parameter values are $Co = 1.2 \times 10^{-7}$ and $St = 0.06$ (cases Flo6-9).}
    \label{fig:PDF_varying_G}
\end{figure}

In the following, we analyze the deformation in time of those flocs that maintain their identity over the time interval $\Delta T$, by keeping track of their characteristic diameter $D_f$. Accordingly, we distinguish between those flocs within the fraction $\theta_{id}$ whose value of $D_f$ increases or stays constant during $\Delta T$, $\theta_{id,gro}$, and those whose diameter decreases, $\theta_{id,shr}$
\begin{equation}  \label{eq:PDF flocs 2}
    \theta_{id} = \theta_{id,gro} + \theta_{id,shr} \ .
\end{equation}
Equations (\ref{eq:PDF flocs 1}) and (\ref{eq:PDF flocs 2}) thus yield
\begin{equation} \label{eq:PDF flocs 3}
    \theta_{br} + \theta_{id,gro} + \theta_{id,shr} + \theta_{ad} = 1 \ .
\end{equation}
For the choice of $\Delta T = 3$, Figure \ref{fig:PDF_Re2e3_Ds5.5e4} displays the evolution of these fractions for the representative case Flo6. Interestingly, we find that $\theta_{id,gro}$ is consistently much larger than $\theta_{id,shr}$, which indicates that of those flocs who maintain their identity during $\Delta T$, many more see their value of $D_f$ increase than decrease. Hence, it is much more common for these flocs to deform from a compact shape to an elongated one than vice versa. This consistent difference between $\theta_{id,gro}$ and $\theta_{id,shr}$ can be maintained only if the elongated flocs eventually break. As a general trend, turbulent stresses thus stretch cohesive flocs before eventually breaking them. This confirms earlier numerical results by \citet{nguyen2014numerical} and \citet{gunkelmann2016influence}, who employed conceptually simpler models with `sticky' cohesive particles and observed that compact flocs have greater strength than elongated ones.

\begin{figure}
    \begin{subfigure}{0.5\textwidth}
    \centering
    \caption{}
    \includegraphics[width=0.9\textwidth]{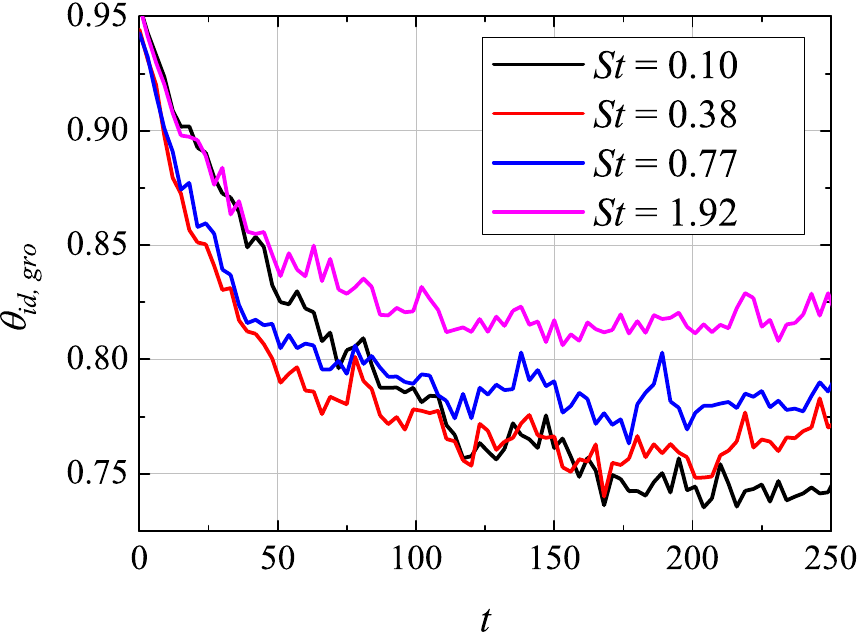}
    \label{fig:theta_mai_gro_varing_St}
    \end{subfigure}
    \begin{subfigure}{0.5\textwidth}
    \centering
    \caption{}
    \includegraphics[width=0.9\textwidth]{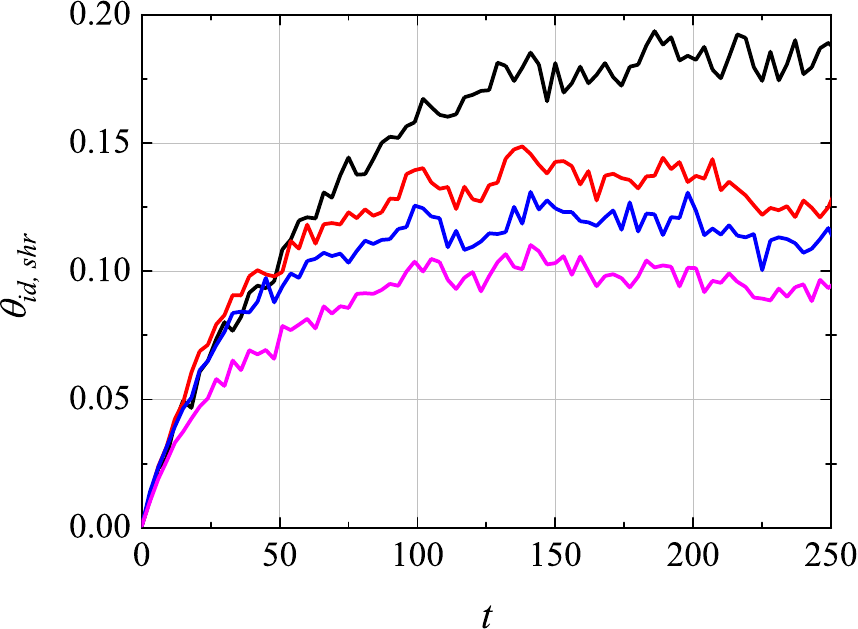}
    \label{fig:theta_mai_shr_varing_St}
    \end{subfigure}
    \caption{Evolution of floc number fractions for different values of $St$. (a) Of those flocs that maintain their identity during $\Delta T$, the fraction $\theta_{id,gro}$ that is stretched increases with $St$; (b) The fraction $\theta_{id,shr}$ whose diameter $D_f$ decreases is reduced for larger $St$. The other parameter values are $Co = 1.2 \times 10^{-7}$ and $G = 0.91$ (cases Flo10-13).}
    \label{fig:PDF_varying_St}
\end{figure}

\begin{figure}
    \begin{subfigure}{0.5\textwidth}
    \centering
    \caption{}
    \includegraphics[width=0.9\textwidth]{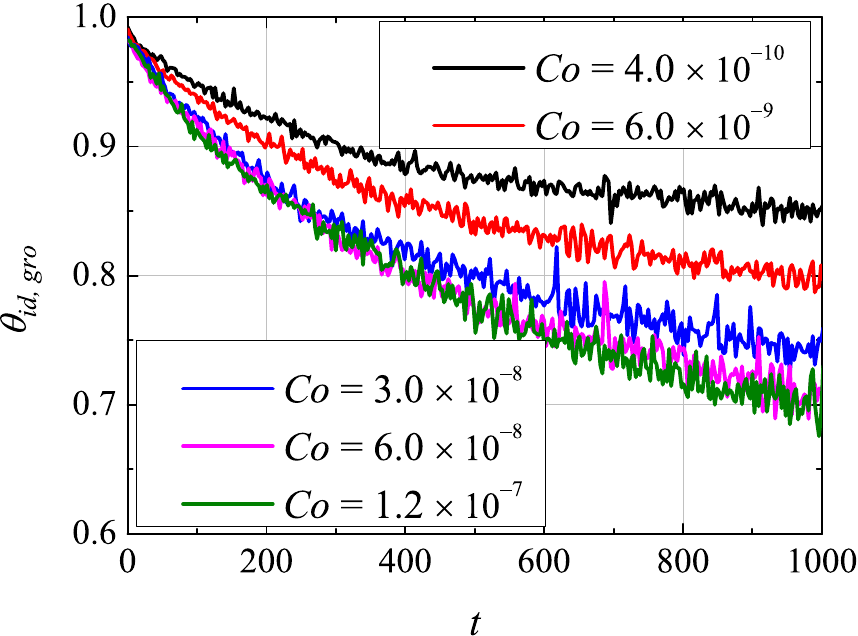}
    \label{fig:theta_mai_gro_varing_Co}
    \end{subfigure}
    \begin{subfigure}{0.5\textwidth}
    \centering
    \caption{}
    \includegraphics[width=0.9\textwidth]{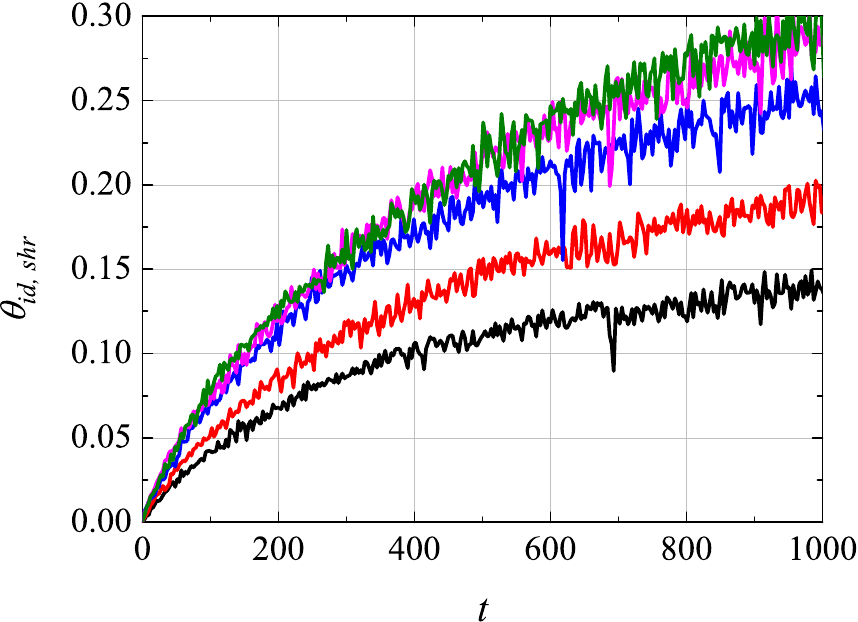}
    \label{fig:theta_mai_shr_varing_Co}
    \end{subfigure}
    \caption{Evolution of floc number fractions for different values of $Co$. (a) Of those flocs that maintain their identity during $\Delta T$, the fraction $\theta_{id,gro}$ that is stretched increases for weaker cohesive forces; (b) The fraction $\theta_{id,shr}$ whose diameter $D_f$ decreases is reduced for weaker cohesive forces. The other parameter values are $St = 0.02$ and $G = 0.29$ (cases Flo1-5).}
    \label{fig:PDF_varying_Co}
\end{figure}

The influence of the shear rate $G$ on the fractions $\theta_{id,gro}$ and $\theta_{id,shr}$ during equilibrium is displayed in Figures \ref{fig:theta_mai_gro_varing_G} and \ref{fig:theta_mai_shr_varing_G}, respectively. For larger values of $G$, the fraction $\theta_{id,gro}$ grows, while $\theta_{id,shr}$ is reduced, which reflects the fact that more intense turbulence tends to elongate the cohesive flocs more strongly. Figures \ref{fig:theta_mai_gro_varing_St} and \ref{fig:theta_mai_shr_varing_St} indicate that larger $St$-values also promote the stretching of those flocs that maintain their integrity, as they increase $\theta_{id,gro}$ and reduce $\theta_{id,shr}$. Figures \ref{fig:theta_mai_gro_varing_Co} and \ref{fig:theta_mai_shr_varing_Co} show that smaller $Co$-values result in the elongation of those flocs that maintain their identity, whereas stronger cohesive forces prompt the flocs to assume a more compact shape.

\subsubsection{Orientation of elongated flocs}

We now investigate the alignment of the elongated flocs with the principal strain directions of the turbulent velocity field. Towards this end, we define an Eulerian fluid velocity difference tensor $\boldsymbol A$ for each floc at time $t$ as 
\begin{equation} \label{eq:defination_nf} 
    {\boldsymbol A}(m,n) = \frac{\boldsymbol u_{f,c}(n) - \boldsymbol u_{f,j}(n)}{\boldsymbol x_{c}(m) - \boldsymbol x_{p,j}(m)} \ ,
\end{equation}
\Rthree{where $m,n = 1, 2, 3$ represent the $x$-, $y$- and $z$-components, respectively,} of the tensor and vectors. ${\boldsymbol x}_c = (x_{c}, y_{c}, z_{c})^{\rm T}$ denotes the location of the floc's center of mass, and the fluid velocity averaged over the volume of the floc is written as ${\boldsymbol u}_{f,c} = \sum_{1}^{N_p}({\boldsymbol u}_{f,i}) / N_p$. The location and fluid velocity at the center of the primary particle $j$ that is located the farthest away from the floc's center of mass are denoted as ${\boldsymbol x}_{p,j} = (x_{p,j}, y_{p,j}, z_{p,j})^{\rm T}$ and ${\boldsymbol u}_{f,j} = (u_{f,j}, v_{f,j} w_{f,j})^{\rm T}$. The orientation of the floc is defined as ${\boldsymbol x}_{f} = {\boldsymbol x}_{p,j} - {\boldsymbol x}_{c}$. Especially for large flocs, ${\boldsymbol x}_c$ and ${\boldsymbol x}_{p,j}$ can be multiple grid spacings apart from each other. \aleck{We remark that $\boldsymbol A$ is defined by sampling the velocity difference at points separated along a line, and it thus represents a simplified approach for considering the influence of the fluid velocity gradients on the whole floc, compared with employing the full coarse-grained velocity gradient tensor \citep{pumir2013tetrahedron}. Hence $\boldsymbol A$ differs from the standard, locally evaluated fluid velocity gradient tensor \citep{ashurst1987alignment, pumir2011orientation, voth2017anisotropic}.} 

We decompose this Eulerian velocity difference tensor $\boldsymbol A = \boldsymbol S + \boldsymbol Q$ into the symmetric velocity difference tensor $\boldsymbol S = {\boldsymbol S}^{\rm T}$, which is similar but not identical to the strain rate tensor, and the anti-symmetric tensor $\boldsymbol Q = -{\boldsymbol Q}^{\rm T}$. The three eigenvalues $r_m$ of the velocity difference tensor $\boldsymbol S$ are ordered as $r_1 > r_2 > r_3$. We remark that the intermediate eigenvalue $r_2$ is automatically zero, by nature of the definition of $\boldsymbol S$. With the three eigenvalues we associate three corresponding orthonormal eigenvectors ${\boldsymbol e}_m$
\begin{equation} \label{eq:defination_nf} 
    \boldsymbol S {\boldsymbol e}_m = r_m {\boldsymbol e}_m \ .
\end{equation}
We define a modified vorticity vector $\boldsymbol \omega = \omega {\boldsymbol e}_{\omega}$ based on the anti-symmetric tensor $\boldsymbol Q$, with magnitude $\omega$ and unit direction vector ${\boldsymbol e}_{\omega}$ \citep{pumir2011orientation}.

We furthermore define a modified deformation gradient tensor $\boldsymbol B$ that characterizes the Lagrangian deformation experienced by a fluid element extending from the floc's center of mass to its primary particle $j$, over the time interval from $t$ to $(t + \Delta t)$, as
\begin{equation} \label{eq:defination_nf} 
    {\boldsymbol B}(m,n) = \frac{\boldsymbol x_{c}(m) - \boldsymbol x_{p,j}(m)}{[\boldsymbol x_{c}(n)+ \Delta t \boldsymbol u_{f,c}(n)] - [\boldsymbol x_{p,j}(n) + \Delta t \boldsymbol u_{f,j}(n)]} \ .
\end{equation}
This modified deformation gradient tensor $\boldsymbol B$ provides a Lagrangian description of the fluid stretching \citep{parsa2011rotation, ni2014alignment}. It differs from the standard locally evaluated deformation gradient tensor, for the same reasons mentioned earlier for the Eulerian velocity difference tensor $\boldsymbol A$.

The Lagrangian stretching tensor $\boldsymbol C = \boldsymbol B \boldsymbol B^T$, obtained from the two symmetric inner products of $\boldsymbol B$ with itself, is similar but not identical to the left Cauchy–Green strain tensor commonly used to define stretching in a Lagrangian basis \citep{chadwick2012continuum}. The three eigenvalues of the Lagrangian stretching tensor $\boldsymbol C$ are ordered as $r_{L1} > r_{L2} > r_{L3}$, and the three corresponding orthonormal eigenvectors are ${\boldsymbol e}_{Lm}$
\begin{equation} \label{eq:defination_nf} 
    \boldsymbol C {\boldsymbol e}_{Lm} = r_{Lm} {\boldsymbol e}_{Lm} \ .
\end{equation}
In the following, we investigate the alignment of ${\boldsymbol x}_{f}$ and ${\boldsymbol e}_{\omega}$ with ${\boldsymbol e}_{m}$ and ${\boldsymbol e}_{Lm}$, respectively.

\begin{figure}
    \begin{subfigure}{0.5\textwidth}
    \centering
    \caption{}
    \includegraphics[width=0.9\textwidth]{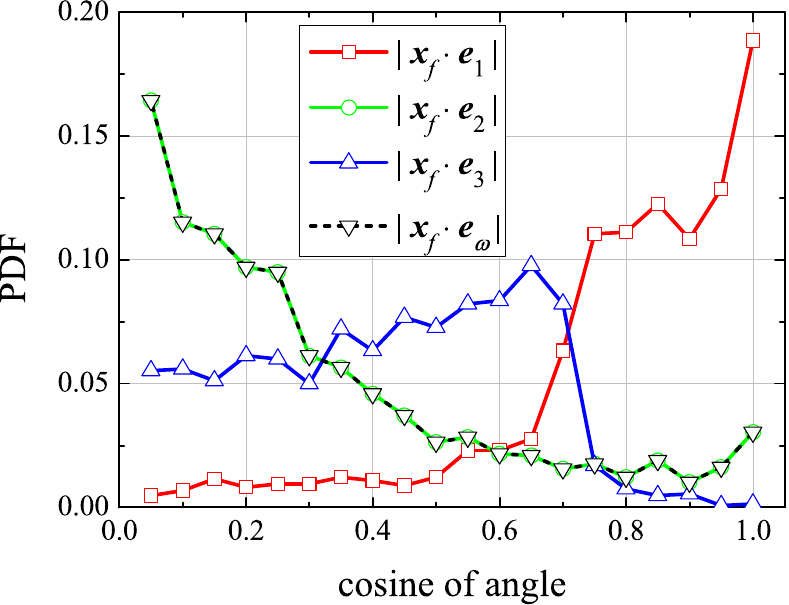}
    \label{fig:Flo9_0.4eta_0.8eta}
    \end{subfigure}
    \begin{subfigure}{0.5\textwidth}
    \centering
    \caption{}
    \includegraphics[width=0.9\textwidth]{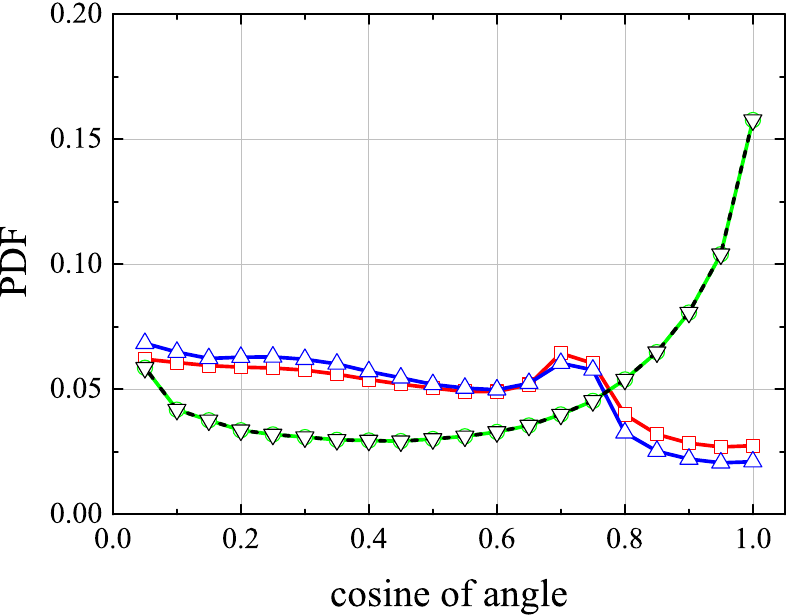}
    \label{fig:Flo9_0.8eta_1.2eta}
    \end{subfigure}
    \begin{subfigure}{0.5\textwidth}
    \centering
    \caption{}
    \includegraphics[width=0.9\textwidth]{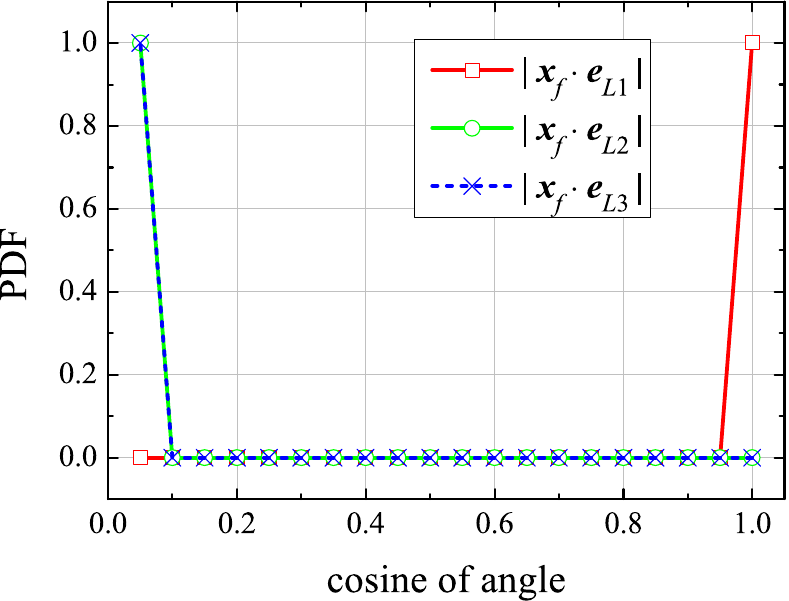}
    \label{fig:Flo9_All_Linear_Flocs_Aligement}
    \end{subfigure}
    \begin{subfigure}{0.5\textwidth}
    \centering
    \caption{}
    \includegraphics[width=0.9\textwidth]{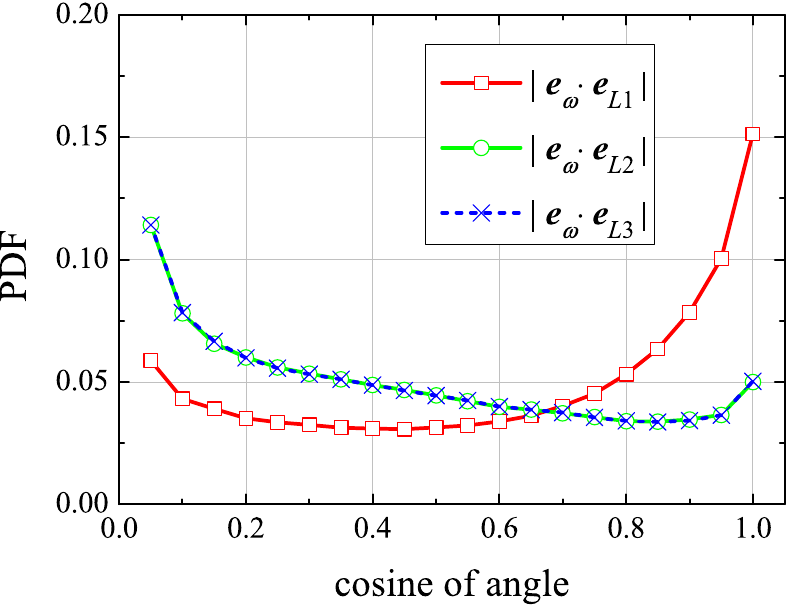}
    \label{fig:Flo9_Vorticity_Aligement}
    \end{subfigure}
    \caption{Floc alignment with the principal directions of the symmetric Eulerian velocity difference tensor for the representative case Flo 9. Results include both the flocculation and the equilibrium stages, for all elongated flocs with $n_f \leqslant 1.2$ and $N_{p,local} \geqslant 2$. The upper two frames show the alignment of the floc orientation ${\boldsymbol x}_f$ with the eigendirections ${\boldsymbol e}_m$ of the symmetric Eulerian velocity difference tensor, and with the vorticity vector ${\boldsymbol e}_{\omega}$: (a) Small flocs with $D_f / \eta < 0.8$; (b) Medium-size flocs with $0.8 \leqslant D_f / \eta \leqslant 1.2$. Small flocs are preferentially aligned with the extensional strain direction, while medium-size flocs tend to align themselves with the intermediate strain direction. The lower two frames show the alignment with the eigendirections ${\boldsymbol e}_{Lm}$ of the Lagrangian deformation tensor: (c) The floc orientation ${\boldsymbol x}_f$; (d) The vorticity vector ${\boldsymbol e}_{\omega}$. Both the flocs and the vorticity vector tend to be aligned with the strongest Lagrangian stretching direction.}
    \label{fig:Alignment of linear flocs}
\end{figure}

We focus on those elongated flocs with $n_f \leqslant 1.2$ and $N_{p} \geqslant 2$, and firstly analyze their alignment with the eigendirections ${\boldsymbol e}_m$ of the Eulerian velocity difference tensor and the vorticity vector ${\boldsymbol e}_{\omega}$ in terms of the magnitude of the angle $\alpha$ between them. We divide the elongated flocs into three different groups, according to the ratio of their characteristic diameter $D_f$ and the Kolmogorov length scale $\eta$. The alignment of small flocs with $D_f/\eta < 0.8$ and medium-size flocs with $0.8 \leqslant D_f/\eta \leqslant 1.2$ is indicated in Figures \ref{fig:Flo9_0.4eta_0.8eta} and \ref{fig:Flo9_0.8eta_1.2eta}, respectively. The alignment of large flocs with $D_f/\eta > 1.2$ is not shown. The results indicate that the modified vorticity vector ${\boldsymbol e}_{\omega}$ is always aligned with the intermediate eigenvector ${\boldsymbol e}_2$, which is consistent with the previous finding by \citet{ashurst1987alignment}. We observe that medium-size flocs are strongly aligned with the intermediate eigenvector ${\boldsymbol e}_2$ and the vorticity vector ${\boldsymbol e}_{\omega}$, as shown in Figure \ref{fig:Flo9_0.8eta_1.2eta}. This result is consistent with previous findings for microscopic axisymmetric rod-like particles in turbulence by \citet{pumir2011orientation}, who noticed that the vortex stretching term ${\boldsymbol A} {\boldsymbol \omega}$ promotes, and the viscous term $\nabla^2 {\boldsymbol \omega} / Re$ opposes, the alignment of ${\boldsymbol x_f}$ with ${\boldsymbol e}_{\omega}$. In contrast, Figure \ref{fig:Flo9_0.4eta_0.8eta} shows that small flocs tend to align themselves with the extensional strain direction ${\boldsymbol e}_1$. For large flocs, we did not observe preferential alignment of the flocs with any of the three eigendirections of the Eulerian velocity difference tensor (not shown).  

The alignment of the elongated flocs and the modified vorticity vector with the eigendirections ${\boldsymbol e}_{Lm}$ of the Lagrangian stretching tensor is shown in Figures \ref{fig:Flo9_All_Linear_Flocs_Aligement} and \ref{fig:Flo9_Vorticity_Aligement}, respectively. The results indicate that the elongated flocs are perfectly aligned, and the modified vorticity vector is strongly aligned with the direction corresponding to the largest eigenvalue ${\boldsymbol e}_{L1}$ of the Lagrangian stretching tensor $\boldsymbol C$. This alignment is consistent with, but even more pronounced than the corresponding previous findings by \Rthree{\citet{parsa2011rotation}} and \citet{ni2014alignment}, due to our definition of the modified deformation gradient tensor $\boldsymbol B$. The perfect alignment of ${\boldsymbol x}_{f}$ with ${\boldsymbol e}_{L1}$ suggests that the present Lagrangian stretching tensor $\boldsymbol C$ is well suited for analyzing the instantaneous alignment of flocs in turbulent flows. 

\subsection{Floc size vs. Kolmogorov length scale} \label{Constraint on flocs}

Several authors have hypothesized that for sufficiently strong turbulence the median floc size should be of the same order as the smallest turbulent eddies \citep{mccave1984size, fettweis2006suspended, coufort2008analysis, kuprenas2018shear}. Others have suggested that even the largest flocs cannot exceed the Kolmogorov length scale \citep{verney2011behaviour}. In the following, we discuss data from the present simulations in order to explore this issue.

\begin{figure}
    \begin{subfigure}{0.5\textwidth}
    \centering
    \caption{}
    \includegraphics[width=0.9\textwidth]{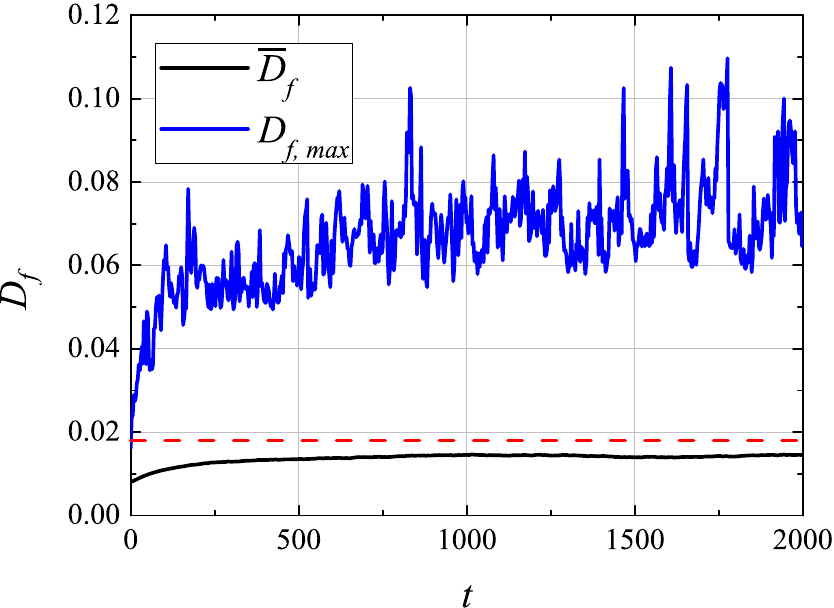}
    \label{fig:Re5e3_Ds5e4_Dfmax}
    \end{subfigure}
    \begin{subfigure}{0.5\textwidth}
    \centering
    \caption{}
    \includegraphics[width=0.9\textwidth]{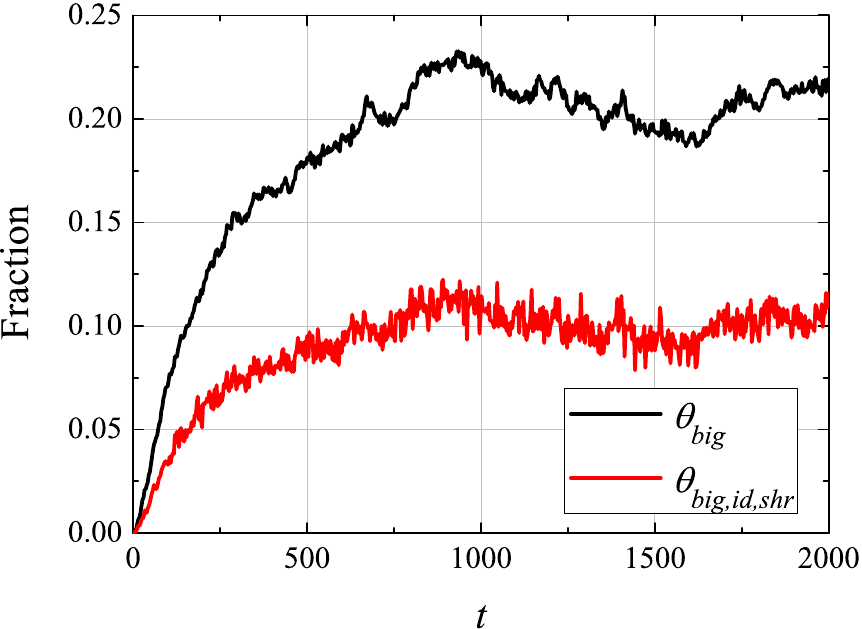}
    \label{fig:Re5e3_Ds5e4_theta_big}
    \end{subfigure}
    \begin{subfigure}{0.5\textwidth}
    \centering
    \caption{}
    \includegraphics[width=0.9\textwidth]{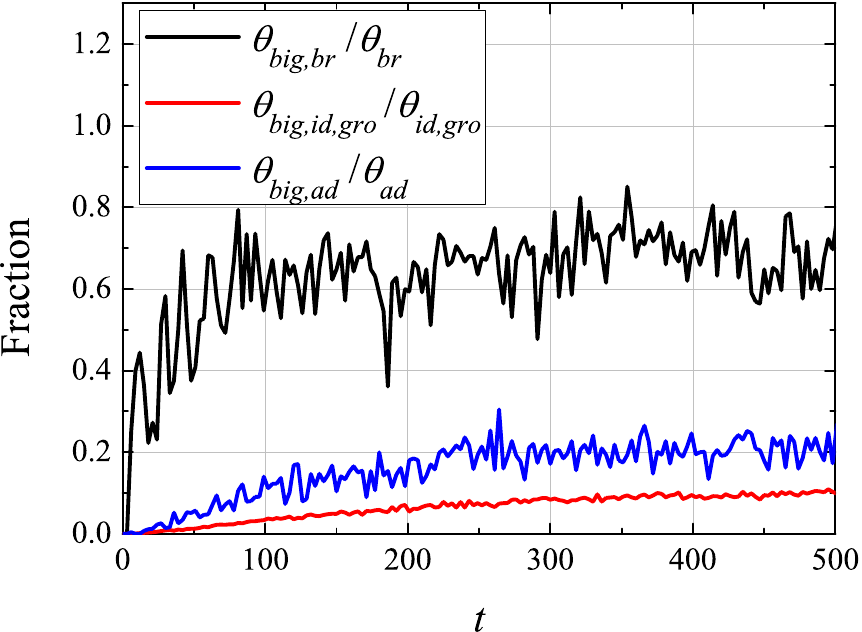}
    \label{fig:Re5e3_Ds5e4_theta_ratio}
    \end{subfigure}
    \begin{subfigure}{0.5\textwidth}
    \centering
    \caption{}
    \includegraphics[width=0.9\textwidth]{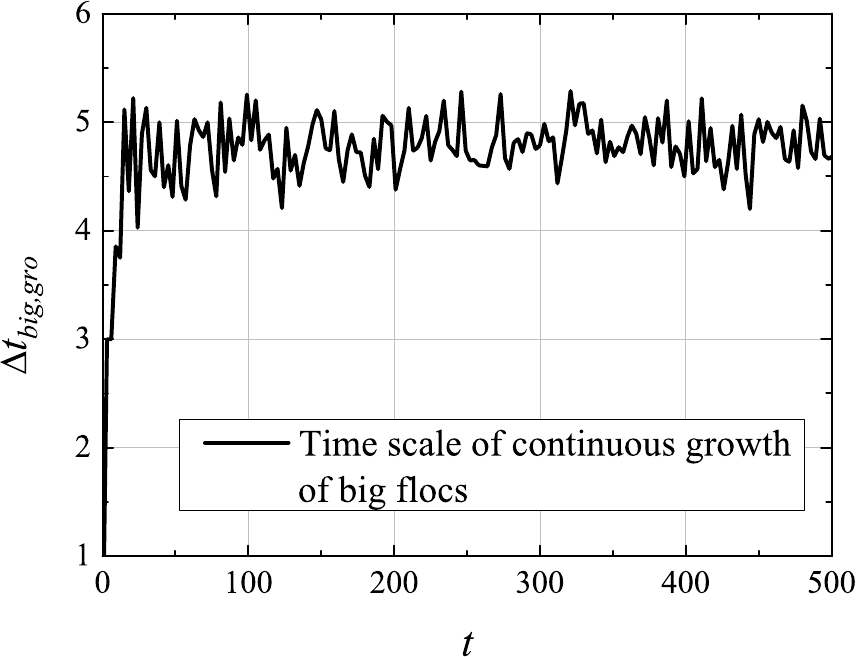}
    \label{fig:Re5e3_Ds5e4_grow_flocs_continue_time}
    \end{subfigure}
    \caption{Constraint on the floc size by the Kolmogorov length scale, for case Flo9 with $\eta/D_p = 2.25$, $G = 0.62$, $St = 0.06$, and $Co = 1.2 \times 10^{-7}$. (a) Temporal evolution of the average and maximum floc diameters, $\overline D_f$ and $D_{f,max}$. The dashed horizontal line indicates the Kolmogorov length scale $\eta$; (b) The fraction $\theta_{big}$ of flocs that are larger than $\eta$, and the fraction $\theta_{big,id,shr}$ of big flocs maintaining their identity that become more compact; (c) The ratios $\theta_{big,br}/\theta_{br}$, $\theta_{big,id,gro}/\theta_{id,gro}$ and $\theta_{big,ad}/\theta_{ad}$; (d) Average time interval $\Delta t_{big,gro}$ over which big flocs exhibit continuous growth.}
    \label{fig: Constraint on flocculation ratio 2.25}
\end{figure}

Figure \ref{fig: Constraint on flocculation ratio 2.25} discusses case Flo9, with $\eta/D_p = 2.25$, $G = 0.62$, $St = 0.06$, and $Co = 1.2 \times 10^{-7}$. Figure \ref{fig:Re5e3_Ds5e4_Dfmax} compares both the average and the maximum floc size to the Kolmogorov scale. It demonstrates that for all times the average floc diameter $\overline D_f$ is smaller than the Kolmogorov length scale $\eta$. However, at any given time the largest floc diameter $D_{f,max}$ is several times larger than $\eta$. We now define "big" flocs as those whose diameter $D_f$ is larger than $\eta$, and we indicate their fraction as
\begin{equation} 
    \theta_{big} = \frac{N_{f,big}}{N_f}  \ ,
\end{equation}
where $N_{f,big}$ is the number of big flocs at a given moment. Analogous to equation (\ref{eq:PDF flocs 3}), we also define the fractions of big flocs that grow, break or maintain their identity, so that we have
\begin{equation} 
    \theta_{big,br} + \theta_{big,id,gro} + \theta_{big,id,shr} + \theta_{big,ad} = \theta_{big}  \ .
\end{equation}
Here the subscripts $br$, $ad$, $id$, $gro$ and $shr$ have the same meanings as in eqn. (\ref{eq:PDF flocs 3}). Figure \ref{fig:Re5e3_Ds5e4_theta_big} demonstrates that $\theta_{big}$ plateaus around a value of 0.2, so that at any given time approximately 20\% of all flocs are larger than the Kolmogorov scale. $\theta_{big,id,shr}$ levels off around 0.1, which indicates that a substantial fraction of these big flocs deform towards a more compact shape while maintaining their identity over $\Delta T = 3$. Figure \ref{fig:Re5e3_Ds5e4_theta_ratio} shows that the ratio $\theta_{big,br}/\theta_{br}$ is stable around 0.6, so that about 60\% of those flocs that break are larger than the Kolmogorov scale $\eta$. The ratio $\theta_{big,id,gro}/\theta_{id,gro}$ levels off around 0.2, meaning that of those flocs that become elongated while maintaining their identity, only about 20\% are big. Hence we can conclude that most of the big flocs tend to either become more compact or to break, but that some continue to grow. This finding is consistent with previous experimental observations by \citet{stricot2010side}, who found that the breakage of big flocs is not instantaneous and depends on the floc strength. 

Figure \ref{fig:Re5e3_Ds5e4_grow_flocs_continue_time} addresses the time scale over which big flocs grow. The duration of the continuous growth of the big flocs is denoted by $\Delta t_{big,gro}$. We remark that $\Delta t_{big,gro}$ is measured for all big flocs until their $D_f$ is smaller than $\eta$. The results indicate that, on average, flocs larger than the Kolmogorov scale keep growing only for the relatively short time period of $\Delta t_{big,gro} \approx 4.8$. This is consistent with previous observations for controls on floc growth in tidal cycle experiments by \citet{braithwaite2012controls}, who found that big flocs cannot resist the turbulent stresses for long, and that they are torn apart quickly. This relatively quick breakage of large flocs in the simulations also agrees with our findings in Section \ref{Deformation of flocs}, which showed that flocs are being continually stretched until they break. 

To summarize, while the size of an individual floc can be larger than the Kolmogorov length for a brief period of time, once $D_f$ becomes bigger than $\eta$, the floc tends to break relatively soon. \Rthree{ Given that the physical parameter ranges listed in table \ref{tab:table1} represent common fluid-particle systems in nature, our simulation data suggest that the average floc size $\overline D_f$ is effectively limited by the Kolmogorov length scale $\eta$ in such systems. We remark, however, that for other classes of primary particles with potentially much stronger bonds it may be possible, in principle, to form flocs that are significantly larger than the Kolmogorov scale.}

\begin{figure}
    \begin{subfigure}{0.5\textwidth}
    \centering
    \caption{}
    \includegraphics[width=0.9\textwidth]{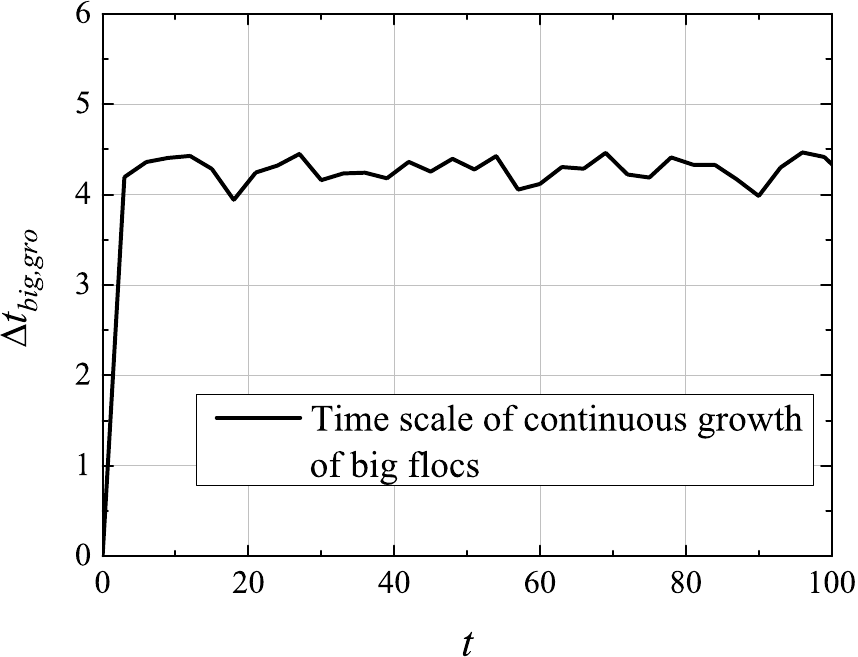}
    \label{fig:Re5e3_Ds1e6_grow_flocs_continue_time}
    \end{subfigure}
    \begin{subfigure}{0.5\textwidth}
    \centering
    \caption{}
    \includegraphics[width=0.9\textwidth]{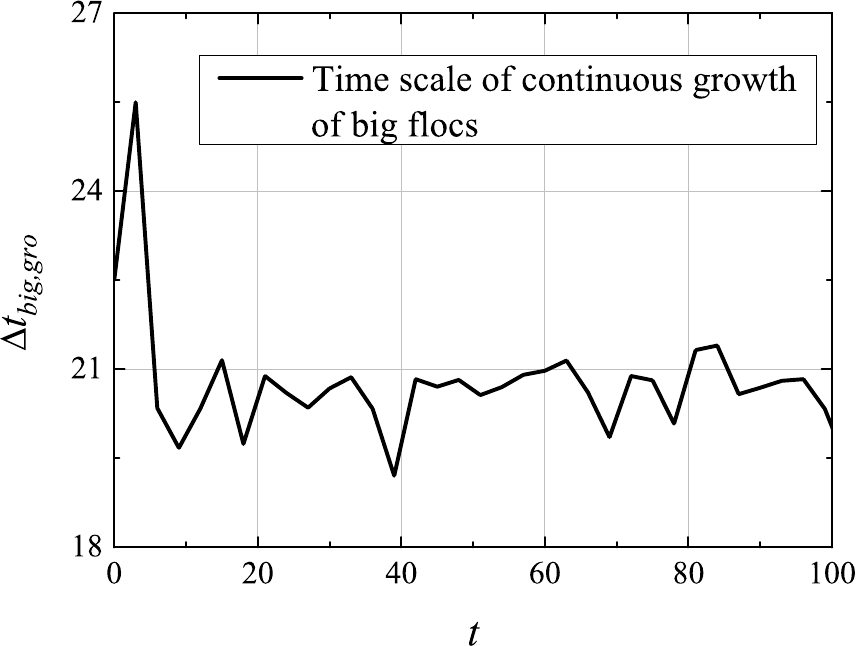}
    \label{fig:Re5e3_Ds1e7_grow_flocs_continue_time}
    \end{subfigure}
    \caption{Average time interval $\Delta t_{big,gro}$ over which big flocs exhibit continuous growth. (a) $\eta/D_p = 1.08$, $Co = 1.2 \times 10^{-7}$, $St = 0.38$, and $G = 2.7$ (case Flo14); (b) $\eta/D_p = 0.65$, $Co = 1.2 \times 10^{-7}$, $St = 1.25$, and $G = 7.4$ (case Flo15).}
    \label{fig: Constraint on flocculation ratio 1.08 and 0.65}
\end{figure}

For cases Flo14 and Flo15, Figure \ref{fig: Constraint on flocculation ratio 1.08 and 0.65} discusses corresponding results regarding the time scale over which big flocs grow. Flo14 employs an increased shear rate $G = 2.7$ along with $\eta/D_p = 1.08$, while Flo15 has $G = 7.4$ and $\eta/D_p = 0.65$. We remark that the ratio $\eta/D_p$ is widely used to classify the primary particles as either `small' if $\eta/D_p > 1$, or as `finite-size' if $\eta/D_p \leqslant 1$ \citep{fiabane2012clustering, costa2015collision, chouippe2015forcing}. Hence Flo14 addresses the small particle scenario, while Flo15 considers finite-size particles. Interestingly, Figure \ref{fig:Re5e3_Ds1e6_grow_flocs_continue_time} shows that the time interval $\Delta t_{big,gro}$ over which big flocs grow for Flo14 is smaller than the corresponding value for Flo9 in Figure \ref{fig:Re5e3_Ds5e4_grow_flocs_continue_time}. This observation indicates that the constraint on floc growth by the turbulent eddies becomes stronger for increasing shear rate $G$, which is consistent with experimental findings for small particles by \citet{braithwaite2012controls}. Those authors had found that the time lag before big flocs break becomes shorter for larger $G$. However, a further increase of the shear rate to $G = 7.4$ in case Flo15, which means that the primary particles now fall into the finite-size category, yields a longer time lag $\Delta t_{big,gro} \approx 20.5$, as shown in Figure \ref{fig:Re5e3_Ds1e7_grow_flocs_continue_time}. While the detailed reasons for this observation will require further investigation, we can conclude that the enhanced control on floc growth by the Kolmogorov length scale for stronger turbulent shear is seen to hold for small primary particles with $\eta/D_p > 1$, although it does not necessarily apply for finite-size primary particles with $\eta/D_p \leqslant 1$.

\section{A new flocculation model with variable fractal dimension} \label{sec:A new flocculation model with variable fractal dimension}

\begin{figure}
    \begin{subfigure}{0.5\textwidth}
    \centering
    \caption{}
    \includegraphics[width=0.9\textwidth]{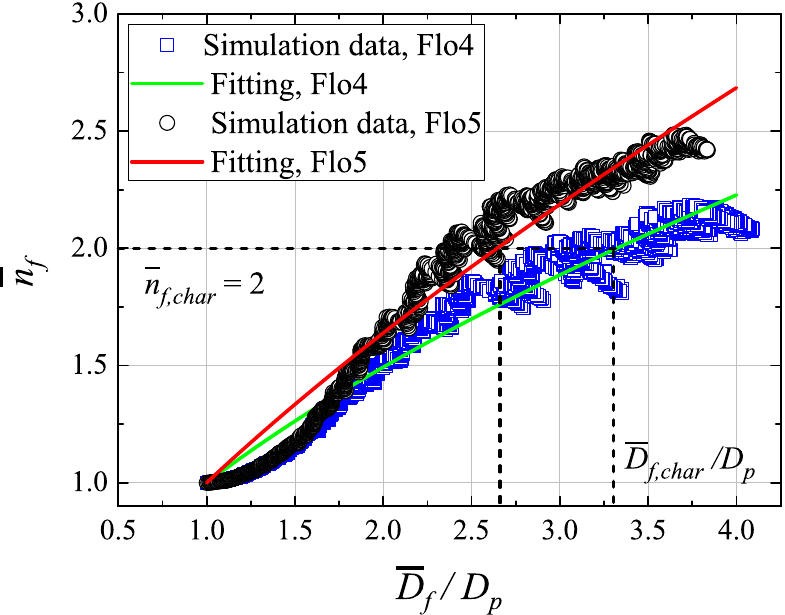}
    \label{fig:typical_fitting_Df_nf}
    \end{subfigure}
    \begin{subfigure}{0.5\textwidth}
    \centering
    \caption{}
    \includegraphics[width=0.9\textwidth]{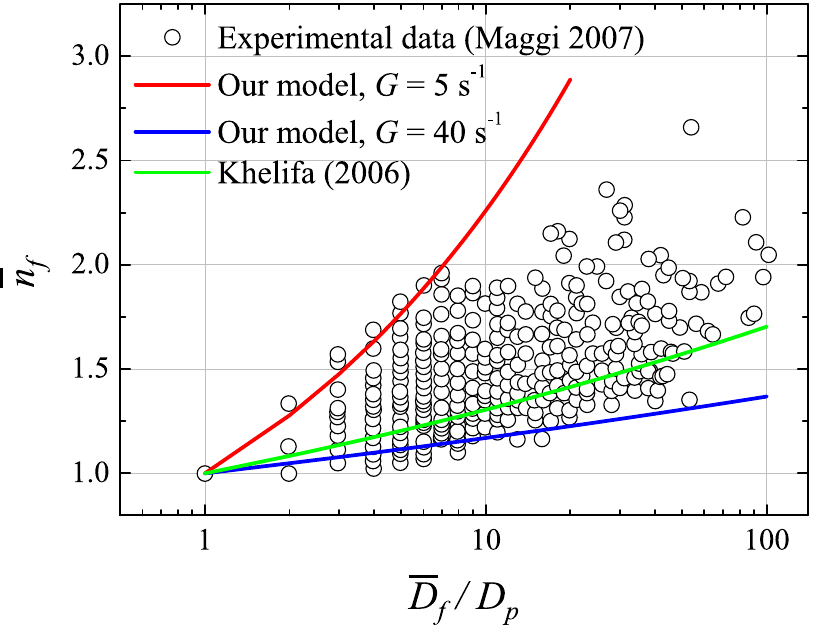}
    \label{fig:nf_models_comparison}
    \end{subfigure}
    \caption{(a) The relationship between the average fractal dimension $\overline n_f$ and the average value $\overline D_f/D_p$, during the flocculation and equilibrium stages. Simulation data and power law fits according to eqn. (\ref{eq:power_nf_Df}) are shown for Flo4 with $Co = 6.0 \times 10^{-8}$, $St = 0.02$, and $G = 0.29$; and for Flo5 with $Co = 1.2 \times 10^{-7}$, $St = 0.02$, and $G = 0.29$; (b) Comparisons between the experimental data of \citet{maggi2007effect}, predictions by the relation of \citet{Khelifa2006II, Khelifa2006I}, and the new relation (\ref{eq:fitting_nf}). The experimental parameters are $D_p = 5 \ \rm{\mu m}$, $\rho_p = 2,650 \ \rm{kg \ m^{-3}}$, $c = 0.5 \ \rm{g \ L^{-1}}$, $\rho_f = 1,000 \ \rm{kg \ m^{-3}}$, $\mu = 0.001 \ \rm{Pa \ s}$ and  $G = 5 \sim 40 \ \rm{ s^{-1}}$. Khelifa's relation (\ref{eq:power_nf_Df}-\ref{eq:k2_Khelifa}) has constant coefficient values $\overline n_{f,char} = 2$, $\overline D_{f,char} = 2,000 \rm{\mu m}$ and updated $k_1 = 1$. The calibration of the empirical coefficient for the new relation (\ref{eq:fitting_nf}) yields $a_3 = 4 \times 10^{-6}$ for $G = 5 \ \rm{ s^{-1}}$, and $a_3 = 4 \times 10^{-5}$ for $G = 40 \ \rm{ s^{-1}}$.}
    \label{fig:relation_Df_nf}
\end{figure}

As indicated by Figures \ref{fig:temporal evolution of floc size G} - \ref{fig:temporal evolution of floc Co}, the average characteristic floc diameter $\overline D_f$ and the average fractal dimension $\overline n_f$ both increase during the flocculation stage, and then remain constant during the equilibrium stage. \Rone{This indicates that flocs of larger size generally have a more compact shape, and that it is difficult for elongated flocs to keep growing in turbulent shear without breaking.} Closer inspection indicates that for all of the cases listed in table \ref{tab:flocculation cases} the relationship between these two quantities can be approximated well by a power law of the form 
\begin{equation} \label{eq:power_nf_Df}
    \overline n_f = k_1 \left( \frac{\overline D_f}{D_p} \right)^{k_2}  \ .
\end{equation}
The condition that $\overline n_f = 1$ for an individual primary particle requires that $k_1 = 1$, while the value of $k_2$ varies as a function of $St$, $Co$ and $G$. Typical fitting results are shown in Figure \ref{fig:typical_fitting_Df_nf} for cases Flo4 and Flo5. This power law relationship allows us to obtain the average fractal dimension $\overline n_f$ during flocculation as a function of the average floc diameter $\overline D_f$, rather than assuming a constant fractal dimension, as was done in earlier investigations \citep{winterwerp1998simple, kuprenas2018shear, zhao2020efficient}.

The power law (\ref{eq:power_nf_Df}) is closely related to the earlier study by \citet{Khelifa2006II, Khelifa2006I}. However, those authors assumed that an individual primary particle has $n_f = 3$, and consequently they set $k_1 = 3$. For the exponent $k_2$ they proposed an empirical correlation of the form
\begin{equation} \label{eq:k2_Khelifa}
    k_2 = \frac{\log(\overline n_{f,char} / k_1)}{\log(\overline D_{f,char}/D_p)}  \ ,
\end{equation}
where $\overline D_{f,char}$ denotes the characteristic floc size that exhibits the characteristic fractal dimension $\overline n_{f,char}$. As a general rule, $\overline D_{f,char}$ and $\overline n_{f,char}$ should be evaluated from experiments before one can then determine $k_2$ from (\ref{eq:k2_Khelifa}). Should that not be feasible, \citet{Khelifa2006II, Khelifa2006I} suggested assuming constant values of $\overline n_{f,char} = 2$ and $\overline D_{f,char} = 2,000 \rm{\mu m}$, which yields a constant value for $k_2$ that depends only on the primary particle size $D_p$. As Figure \ref{fig:typical_fitting_Df_nf} indicates, however, $k_2$ should be a function of $G$, $St$ and $Co$ even for a constant $D_p$, since $\overline n_{f,char} = 2$ is associated with different average floc sizes $\overline D_{f,char}/D_p$ in cases Flo4 and Flo5. Hence, even though equation (\ref{eq:k2_Khelifa}) has been widely used to describe the fractal dimension of flocs \citep{maggi2007effect, son2009effect, klassen2017three}, we will now try to refine this scaling law by accounting for the dependence of $k_2$ on $St$, $Co$ and $G$.

By fitting the simulation results for all of the cases Flo1 - 15, we obtain a relationship for $k_2$ of the form
\begin{equation} \label{eq:fitting_k2}
    k_2 = 0.44St^{-0.018}Co^{0.096}G^{-1.5}  \ ,
\end{equation}
with an R-squared value of 0.97. We remark that in a laboratory experiment or field investigation it may be challenging to evaluate the Stokes number $St$ as defined in equation (\ref{eq:defination_St}), if the $rms$-velocity $u_{rms}$ is unknown. To overcome this difficulty, we follow the approach taken in our earlier work \citep{zhao2020efficient}, where we defined the characteristic Stokes number $St_{char}$ and cohesive number $Co_{char}$ by employing the characteristic fluid velocity $u_{char} = 0.25(G/Re)^{0.5}$ instead of $u_{rms}$, so that
\begin{equation} \label{eq:St_char}
    St_{char} = \frac{St \, u_{char}}{u_{rms}} = \frac{\rho_s D_p^2 \, u_{char} Re}{18 \eta} \ ,
\end{equation}
\begin{equation} \label{eq:Co_char}
    Co_{char} = \frac{Co}{\eta^2 \, u_{char}^2} \ .
\end{equation}
Here $Re$ and $Co$ are of the form defined in (\ref{eq:dimensionless fluid momentum conservation}) and (\ref{eq:Co_definition}), respectively. Note that $u_{char}$ and $\eta$ in equations (\ref{eq:St_char}) and (\ref{eq:Co_char}) are dimensionless. Based on $St_{char}$ and $Co_{char}$, a fit of the simulation data yields the relationship for $k_2$
\begin{equation} \label{eq:fitting_k2_char}
    k_2 = \frac{St_{char}^{-1.9}Co_{char}^{0.1}}{1.3 \times 10^5}  \ ,
\end{equation}
which has an R-squared value of 0.86. \Rone{Here $St_{char}$ captures the strongly inverse influence of the shear rate $G$ on $k_2$.} By substituting (\ref{eq:fitting_k2_char}) into (\ref{eq:power_nf_Df}), we obtain a new model for the average fractal dimension $\overline n_f$ of the form
\begin{equation} \label{eq:fitting_nf}
    \overline n_f = \left(\frac{\overline D_f}{D_p}\right)^{a_3 \, St_{char}^{-1.9}Co_{char}^{0.1}/(1.3 \times 10^5)}\ .
\end{equation}
For the specific range of physical parameters listed in table \ref{tab:table1}, $a_3=1$ yields optimal agreement with a maximum deviation of $\pm 30\%$ from the simulation data. As we will see below, this value of $a_3$ is not universally optimal, so that $a_3$ will have to be recalibrated for other parameter ranges. In the following, we will compare predictions for the fractal dimension by the new relation (\ref{eq:fitting_nf}) with corresponding ones by the earlier relation of Khelifa \& Hill (\ref{eq:power_nf_Df}-\ref{eq:k2_Khelifa}). 

Employing the approach of \citet{maggi2004method}, \citet{maggi2007effect} estimate the time evolution of the average fractal dimension $\overline n_f$ and floc size $\overline D_f$ in experiments with constant turbulent shear rates $G = 5, 10, 20$ and 40 $\rm{s^{-1}}$, respectively. The suspended cohesive sediment in the experiments has a primary particle diameter $D_p = 5 \ \rm{\mu m}$, density $\rho_p = 2,650 \ \rm{kg \ m^{-3}}$, and concentration $c = 0.5 \ \rm{g \ L^{-1}}$. Since the authors assume $\overline n_f = 3$ for flocs with one particle while we set $\overline n_f = 1$ for that situation, we have to convert their original experimental data before we can compare them with the present simulation results. The details of the conversion are discussed in appendix \ref{app: Conversion of the experimental data}, and the converted data are presented in Figure \ref{fig:nf_models_comparison}. In addition, we set the dimensional Kolmogorov length $\eta = [\mu / (\rho_f G)]^{0.5} \rm{m}$ and the Hamaker constant $A_H = 1.0 \times 10^{-20} \ \rm{J}$ to obtain the characteristic values $St_{char}$ and $Co_{char}$ according to (\ref{eq:St_char})-(\ref{eq:Co_char}). Since the experimental shear rates $G = 5 \sim 40 \ \rm{ s^{-1}}$ are much smaller than the simulation values $G = 3.7 \times 10^3 \sim 9.5 \times 10^4 \ \rm{ s^{-1}}$, we have to recalibrate the constant $a_3$ required for our model (\ref{eq:fitting_nf}) from the experimental data. Based on the fact that the exponent $k_2$ should decrease for increasing $G$, we obtain $a_3 = 4 \times 10^{-6}$ for the minimum experimental shear rate $G = 5 \rm{ s^{-1}}$, and $a_3 = 4 \times 10^{-5}$ for the maximum experimental shear rate $G = 40 \rm{ s^{-1}}$, respectively. Figure \ref{fig:nf_models_comparison} demonstrates that the present relation successfully reproduces the range of experimental data for different $G$-values, whereas Khelifa \& Hill's relation does not account for variations in $G$. At the same time, we do need to keep in mind that the present model does require a recalibration of $a_3$ for different experimental parameter ranges.

In order to develop a variable fractal dimension model for the transient stages, we build on the approach taken in our recent investigation \citep{zhao2020efficient}. There we conducted cohesive sediment simulations for a steady, two-dimensional cellular flow model. Based on the simulation data, we proposed an analytical flocculation model of the form
\begin{subequations} \label{eq:new model with a constant fractal dimension}
    \begin{eqnarray} 
        \label{eq:Our new model Df}
        \overline D_f & = & (\overline N_{p}) ^ {\frac{1}{\overline n_f}} D_p \ , \\
        \label{eq:Our new model Np,local}
        \overline N_p & = & \frac{1}{(1 / \overline N_{p,in} - 1 / \overline N_{p,eq}) e^{b t} +  1 / \overline N_{p,eq}} \ , 
        \\
        \label{eq:Our new model Np,local,max}
        \overline N_{p,eq} & = & \left\{
            \begin{array}{ll}
                N,  \quad  if \ \overline N_{p,eq} \geqslant N \ , \\ 
                \\
                8.5 a_1 St_{char}^{0.65} Co_{char}^{0.58} D_{p,char}^{-2.9} \phi_p^{0.39} \rho_s^{-0.49} (W+1)^{-0.38} \ , \quad otherwise  \ ,
            \end{array}\right.
        \\
        \label{eq:Our new model b}
        b & = & \left\{
            \begin{array}{ll}
                -0.7 a_2 St_{char}^{0.36} Co_{char}^{-0.017} D_{p,char}^{-0.36} \phi_p^{0.75} \rho_s^{-0.11} (W+1)^{-1.4},  \quad St_{char} \leqslant 0.7  \ , \\
                \\
                -0.3 a_2 St_{char}^{-0.38} Co_{char}^{0.0022} D_{p,char}^{-0.61} \phi_p^{0.67} \rho_s^{0.033} (W+1)^{-0.46}, \quad St_{char} > 0.7  \ .
            \end{array}\right.
    \end{eqnarray}
\end{subequations}
Here $\overline N_{p,in}$ and $\overline N_{p,eq} = N/N_{f,eq}$ indicate the average number of primary particles per floc at the initial time and during the equilibrium stage, respectively. \Rone{$|b|$ denotes the rate of change in the number of flocs, where a bigger $|b|$ indicates a faster increase of the mean number of primary particles per floc $\overline N_{p}$ during flocculation.} $D_{p,char} = D_p/\eta$ is the characteristic primary particle diameter, and $W$ represents the Stokes settling velocity. $a_1$ and $a_2$ are empirical coefficients that need to be calibrated via comparison with experiments or simulations. Under the assumption of a constant average fractal dimension $\overline n_f = 2$, and for given values of $N$, $\overline N_{p,in}$, $St_{char}$, $Co_{char}$, $D_{p,char}$, $\phi_p$, $\rho_s$ and $W$, this model predicts the transient floc size $\overline D_f$ and the average number of particles per floc $\overline N_{p}$ as functions of time. Model results were presented in \citet{zhao2020efficient}. As the present simulations show, however, assuming a constant average fractal dimension represents a serious limitation, cf. Figures \ref{fig:nnf_more_t_varying_G}, \ref{fig:nnf_more_t_varying_St} and \ref{fig:nnf_t_varying_Co}, which we aim to overcome in the following.

Towards this end, we combine equations (\ref{eq:fitting_nf}) and (\ref{eq:new model with a constant fractal dimension}) to obtain a new flocculation model (termed the `present model') that allows for a variable fractal dimension. This model yields predictions of the floc size $\overline D_f$, the number of particles per floc $\overline N_{p}$, and the fractal dimension $\overline n_f$ as functions of time. Since equations (\ref{eq:fitting_nf}) and (\ref{eq:Our new model Df}) need to be solved concurrently, the model cannot be written in closed form. However, due to the narrow range of the average fractal dimension $1 \leqslant \overline n_f \leqslant 3$, an iterative solution can easily be obtained. 

In analogous fashion, we can link the variable fractal dimension relation (\ref{eq:power_nf_Df}) - (\ref{eq:k2_Khelifa}) by \citet{Khelifa2006II, Khelifa2006I} to our previous flocculation model (\ref{eq:new model with a constant fractal dimension}), to obtain the `combined model.' A list of all models discussed here is provided in table \ref{tab:model list} for convenience. We now proceed to assess their performance.

\begin{table}
  \begin{center}
\def~{\hphantom{0}}
  \begin{tabular}{ccc}
     Terminology \ \ \ & Composition of models \ \ \ & Predict the evolution of   \\ 
    \\
    \citet{Khelifa2006II, Khelifa2006I} relation \ \  & (\ref{eq:power_nf_Df}) and (\ref{eq:k2_Khelifa})  & average fractal dimension $\overline n_f(\overline D_f)$     \\ 
    \\
    New fractal relation \ \  & (\ref{eq:fitting_nf})  & average fractal dimension $\overline n_f(\overline D_f)$     \\ 
    \\
    \citet{zhao2020efficient} model \ \  & (\ref{eq:new model with a constant fractal dimension})  & average floc size $\overline D_f(t)$     \\ 
    \\
    Present model \ \  &  (\ref{eq:fitting_nf}) and (\ref{eq:new model with a constant fractal dimension})   & both $\overline D_f(t)$ and $\overline n_f(t)$  \\ 
    \\
    Combined model \ \  & (\ref{eq:power_nf_Df}), (\ref{eq:k2_Khelifa}) and (\ref{eq:new model with a constant fractal dimension})  & both $\overline D_f(t)$ and $\overline n_f(t)$  \\ 
    \\
  \end{tabular}
  \caption{Typical models cited, proposed and implemented in the present work.}
  \label{tab:model list}
  \end{center}
\end{table}

\begin{figure}
    \begin{subfigure}{0.5\textwidth}
    \centering
    \caption{}
    \includegraphics[width=0.9\textwidth]{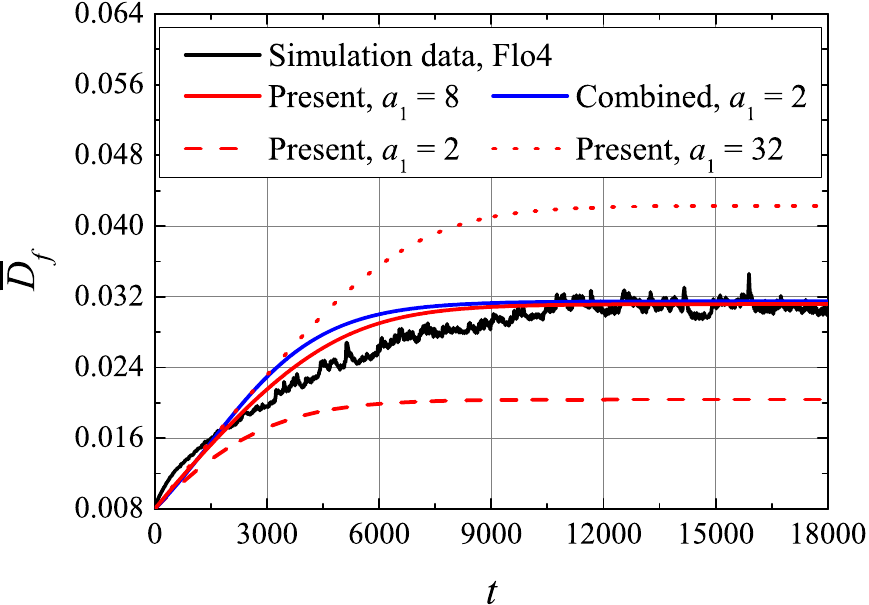}
    \label{fig:Comparsion_Df}
    \end{subfigure}
    \begin{subfigure}{0.5\textwidth}
    \centering
    \caption{}
    \includegraphics[width=0.9\textwidth]{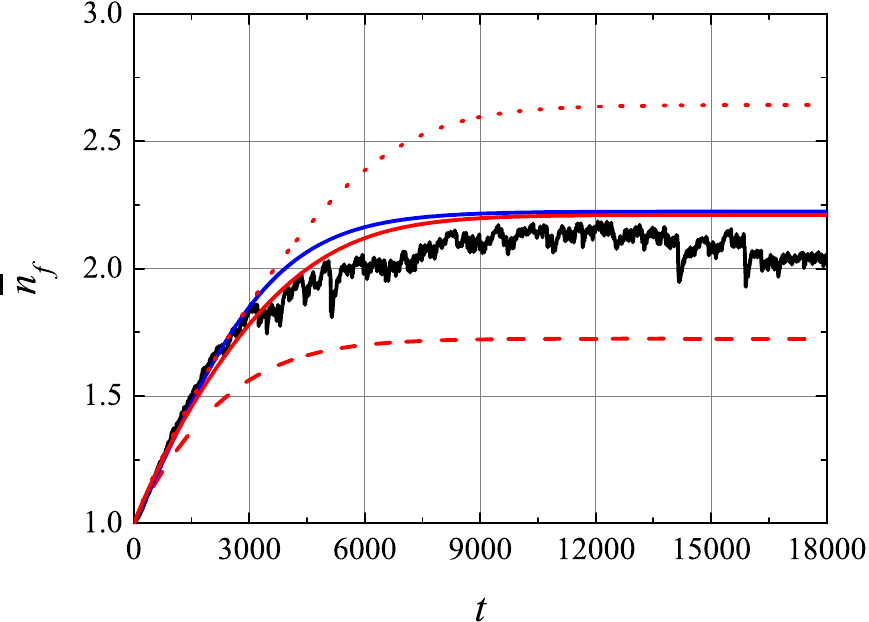}
    \label{fig:Comparsion_nf}
    \end{subfigure}
    \caption{ Comparisons between the numerical data and predictions by the present model and the combined model listed in table \ref{tab:model list}, simulation data of case Flo4 with $Co = 6.0 \times 10^{-8}$, $St = 0.02$ and $G = 0.29$ is selected. (a) Calibration predictions for the temporal evolution of the average floc izes $\overline D_f$, the calibrated coefficients in the models are  $a_2 = 0.5$ and $a_3 = 1$, the constant $k_1 = 1$; (b) Comparisons for the temporal evolution of the average fractal dimension $\overline n_f$.}
    \label{fig:new_model_comparsion}
\end{figure}

By calibrating with the average floc size data for simulation Flo4, we determine the empirical coefficients for the `present model' as $a_1 = 8$, $a_2 = 0.5$ and $a_3 = 1$, shown as solid red line in Figure \ref{fig:Comparsion_Df}. We then employ the present model to predict the average fractal dimension for Flo4 as function of time. Figure \ref{fig:Comparsion_nf} indicates good agreement between the predictions and the simulation data. In complete analogy, we determine the empirical coefficients for the `combined model' as $a_1=2$, $a_2 = 0.5$ and $k_1 = 1$, which yields the solid blue line in Figure \ref{fig:Comparsion_Df}. The average fractal dimension $\overline n_f$ predicted by the combined model is very close to that of the present model and to the simulation data, which suggests that both models are able to predict the average fractal dimension quite accurately.

In applications, it may be difficult to obtain precise calibration values for $a_1$, so that it is important to establish the robustness of the present model with regard to uncertainties in the value of $a_1$. In order to assess this robustness, we ran the present model for $a_1=2$ and $a_1=32$, instead of the optimal value $a_1=8$ that we had obtained earlier from the calibration. The results, shown in Figure \ref{fig:new_model_comparsion} as dashed and dotted red lines, indicate that the model predictions are reasonably robust with regard to uncertainties in the value of $a_1$.

To summarize, our new fractal relation (\ref{eq:fitting_nf}) no longer has the limitation associated with assuming a constant value for $k_2$ in equations (\ref{eq:power_nf_Df}) and (\ref{eq:k2_Khelifa}) when predicting the variable fractal dimension $\overline n_f$. In addition, we observe that predictions of the floc size $\overline D_f$ and the fractal dimension $\overline n_f$ as functions of time by the present flocculation model (\ref{eq:fitting_nf}) and (\ref{eq:new model with a constant fractal dimension}) are fairly robust with respect to uncertainties that arise when calibrating the empirical coefficients by means of experimental data.

\section{Conclusions} \label{sec:Conclusions}

In the present investigation we have employed one-way coupled simulations to explore the dynamics of cohesive particles in homogeneous isotropic turbulence. The simulations account for the Stokes drag, as well as lubrication, cohesive and direct contact forces. They demonstrate the existence of a transient flocculation phase which is characterized by the growth of the average floc size. This flocculation phase is followed by a statistically steady equilibrium phase governed by a balance between floc growth and breakup. The simulations provide information about the temporal evolution of the floc size and shape, as a result of aggregation, breakage and deformation, and as function of the governing parameters. In general, we find that larger turbulent shear and weaker cohesive forces limit the floc size and result in elongated floc shapes. Flocculation proceeds most rapidly during the transient stage when the Stokes number of the primary particles based on the Kolmogorov scales is of order unity. During the transient stage cohesive forces of intermediate strength yield the largest flocs. On one hand, these intermediate cohesive forces are strong enough to result in the rapid aggregation of primary particles, but on the other hand they are not so strong as to pull them into a compact shape. During the equilibrium stage, stronger cohesive forces produce larger flocs. Small Stokes numbers and weak turbulence typically lead to a later onset of the equilibrium stage. The equilibrium floc size distribution exhibits a preferred size as function of the cohesive number. This distribution decays exponentially for larger floc sizes. The simulation results indicate that flocs are generally elongated by turbulent stress before they eventually break. We observe that flocs close to the Kolmogorov scale in size preferentially align themselves with the intermediate strain direction and the vorticity vector. Flocs that are smaller than the Kolmogorov scale, on the other hand, tend to align themselves with the direction of extensional strain. The simulation results furthermore demonstrate that flocs generally align themselves with the strongest Lagrangian stretching direction. The simulations show that the average floc size is effectively limited by the Kolmogorov scale, and can at most exceed it marginally. However, individual flocs can grow larger than the Kolmogorov scale for a limited amount of time. Based on the simulation data we propose a novel flocculation model that allows for a variable fractal dimension, which enables us to predict the temporal evolution of the floc size and shape, as a function of the governing dimensionless parameters, after some limited calibration. Predictions by the new model are fairly robust and cover a broad range of parameters.

\vspace{.4in}
\noindent
{\bf Acknowledgements}\\
EM gratefully acknowledges support through NSF grants CBET-1803380 and OCE-1924655, as well as by the Army Research Office through grant W911NF-18-1-0379. TJH received support through NSF grant OCE-1924532. KZ is supported by the National Natural Science Foundation of China through the Basic Science Center Program for Ordered Energy Conversion through grant 51888103, as well as by the China Scholarship Council. BV gratefully acknowledges support through German Research Foundation (DFG) grant VO2413/2-1. Computational resources for this work used the Extreme Science and Engineering Discovery Environment (XSEDE), which is supported by NSF grant TG-CTS150053.

\vspace{.2in}
\noindent
{\bf Declaration of Interests}\\
The authors report no conflict of interest.

\appendix
\section{Conversion of the experimental data} \label{app: Conversion of the experimental data}

\begin{figure}
    \centering
    \includegraphics[width=0.55\textwidth]{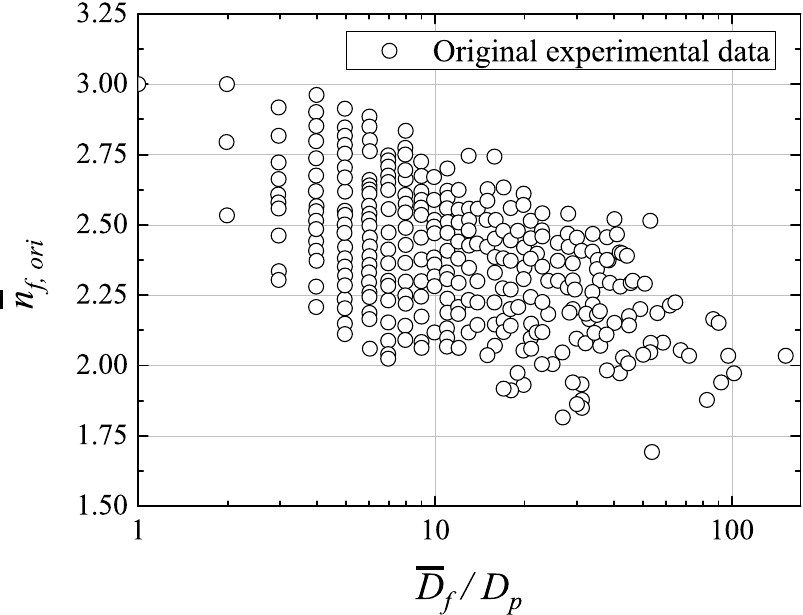}
    \caption{ Original experimental data of \citet{maggi2007effect} for the experimental parameter values $D_p = 5 \ \rm{\mu m}$, $\rho_p = 2650 \ \rm{kg \ m^{-3}}$, $c = 0.5 \ \rm{g \ L^{-1}}$, $\rho_f = 1,000 \ \rm{kg \ m^{-3}}$, $\mu = 0.001 \ \rm{Pa \ s}$ and  $G = 5 \sim 40 \ \rm{ s^{-1}}$.}
    \label{fig:original_experimental_data_Maggi}
\end{figure}

\citet{maggi2007effect} measured the floc size and evaluated the fractal dimension $\overline n_{f,ori}$ in experiments by setting the fractal dimension of an individual primary particle to three. Taking their experimentally measured floc size as the characteristic floc diameter $D_f$ in (\ref{eq:defination_Df}), the original experimental data are shown in Figure \ref{fig:original_experimental_data_Maggi}. For each pair of $\overline n_{f,ori}$ and $\overline D_f/D_p$, we can obtain the average floc size $\overline D_{f,char}$ as
\begin{equation} \label{eq:conversion_Df,char}
    \overline D_{f,char} = D_p10^{[\log(\overline n_{f,char} / k_{1,ori})]/k_{2,ori}}  \ ,
\end{equation}
where the characteristic fractal dimension $\overline n_{f,char} = 2$, the diameter of primary particles $D_p = 5 \ \rm{\mu m}$, $k_{1,ori} = 3$, and
\begin{equation} \label{eq:conversion_k2,ori}
    k_{2,ori} = \frac{\log(\overline n_{f,ori}/k_{1,ori})}{\log(\overline D_{f}/D_p)}  \ .
\end{equation}
The converted fractal dimension $\overline n_{f}$ for the corresponding $\overline D_f/D_p$ in the experiments can then be obtained from
\begin{equation} \label{eq:conversion_nf}
    \overline n_{f} = k_{1}(\frac{\overline D_{f}}{D_p})^{k_2}  \ ,
\end{equation}
where $k_1 = 1$ and
\begin{equation} \label{eq:conversion_k2}
    k_{2} = \frac{\log(\overline n_{f,char}/k_{1})}{\log(\overline D_{f,char}/D_p)}  \ .
\end{equation}
The converted experimental data are shown in Figure \ref{fig:nf_models_comparison}.

\bibliographystyle{jfm}
\bibliography{main_document}

\end{document}